\documentclass[%
 twocolumn,
 amsmath,amssymb,
 aps,
 floatfix,
]{revtex4-2}

\usepackage{amssymb,amsmath,amstext}
\usepackage{graphicx}
\usepackage{bm}
\usepackage{hyperref}
\usepackage{xcolor}

\usepackage{physics}
\usepackage{ulem}

\begin{document}

\title{Splay-induced order in systems of hard wedges}

\author{Piotr Kubala}
\affiliation{Institute of Theoretical Physics, Jagiellonian University in Krak\'ow, \L{}ojasiewicza 11, 30-348 Krak\'ow, Poland }
\author{Micha\l{} Cie\'sla}
\affiliation{Institute of Theoretical Physics, Jagiellonian University in Krak\'ow, \L{}ojasiewicza 11, 30-348 Krak\'ow, Poland }
\author{Lech Longa}
\affiliation{Institute of Theoretical Physics, Jagiellonian University in Krak\'ow, \L{}ojasiewicza 11, 30-348 Krak\'ow, Poland }


\begin{abstract}
The main objective of this work is to clarify the role that wedge-shaped elongated 
molecules, \textit{i.e.}, molecules 
with one end wider than the other, can play in stabilizing orientational order. 
The focus is exclusively on entropy-driven self-organization induced 
by purely excluded volume interactions. 
Drawing an analogy to RM734 (4-[(4-nitrophe-noxy)carbonyl]phenyl2,4-dimethoxybenzoate), 
which is known to stabilize ferroelectric nematic ($\text{N}_\text{F}$)
and nematic splay ($\text{N}_\text{S}$) phases, 
and assuming that molecular biaxiality is of secondary importance, we consider monodisperse systems 
composed of hard molecules. Each molecule is modeled using six colinear tangent spheres 
with linearly decreasing diameters. Through hard-particle, constant-pressure 
Monte Carlo simulations, 
we study the emergent phases as functions of the ratio between the smallest and largest diameters
of the spheres (denoted as $d$) and the packing fraction ($\eta$). To analyze global and local 
molecular orderings, we examine molecular configurations in terms of nematic, smectic, 
and hexatic order parameters.
Additionally, we investigate the radial pair distribution function, 
polarization correlation function,
and the histogram of angles between molecular axes. The latter characteristic is 
utilized to quantify local splay. 
The findings reveal that splay-induced deformations drive unusual 
long-range orientational order at relatively high packing fractions ($\eta > 0.5$), 
corresponding to crystalline phases.
When $\eta < 0.5$, only short-range order is affected, and in addition 
to the isotropic liquid, only the standard nematic and smectic A liquid crystalline 
phases are stabilized. 
However, for $\eta > 0.5$, apart from the ordinary non-polar hexagonal crystal, 
three new frustrated crystalline polar blue phases with 
long-range splay modulation are observed: 
antiferroelectric splay crystal ($\text{Cr}_\text{S}\text{P}_\text{A}$), 
antiferroelectric double splay crystal ($\text{Cr}_\text{DS}\text{P}_\text{A}$) 
and ferroelectric double splay crystal ($\text{Cr}_\text{DS}\text{P}_\text{F}$).
Finally, we employ Onsager-Parsons-Lee Local Density Functional Theory to investigate
whether any sterically-induced (anti-)ferroelectric nematic or smectic-A type of ordering 
is possible for our system, at least in a metastable regime.  
\end{abstract}

\maketitle

\section{Introduction}

The most spectacular discovery of the last decade in the field of liquid crystal
research has been the identification of the ferroelectric nematic ($\text{N}_\text{F}$)
\cite{PhysRevLett.124.037801,chen2020first,Mandle2021,Mandle2022}, 
antiferroelectric nematic splay ($\text{N}_\text{S}$) 
\cite{PhysRevX.8.041025}, and nematic twist-bend ($\text{N}_\text{TB}$) 
\cite{Cestari2011,Borshch2013,Chen2013} phases, all
characterized by various forms of long-range polar order.

Based on current experimental observations, it seems that 
the stabilization of these new nematic phases is linked to 
the strong softening (i.e., reduction by one 
to two orders of magnitude) of one of the Frank elastic constants, 
$K_{ii}$ [9], in the parent, uniaxial nematic ($\text{N}$) phase, 
near the transition to one of the polar nematics.
$K_{ii}$s ($i=1,2,3$) weight here elementary deformations of 
splay ($K_{11}$), twist ($K_{22}$), and bend ($K_{33}$) types
of the undistorted, reference uniaxial nematic state ($\text{N}$)
in the Oseen-Zocher-Frank free energy \cite{Oseen1933,Zocher1933,Frank1958}
\begin{align}
\label{eq:ozf-energy}
    \mathcal{F} &= \frac{1}{2V} \int_V \left[ K_{11} [\vu{n}(\div{\vu{n}})]^2 + 
     K_{22} [\vu{n}\vdot(\curl{\vu{n}})]^2 \right] + \notag\\
    &\hspace{1.6cm}+ \left. K_{33} [\vu{n}\cp(\curl{\vu{n}})]^2\right]; 
\end{align}
$\vu{n}$ denotes the locally preferred orientation of molecules 
referred to as the director, and $V$ is the system's volume. 

While each Frank elastic constant is typically positive and 
on the order of 10 pN, the observed  softening of $K_{ii}$, 
suggests that the lack of 
orientational modulation of $N$ 
is no longer energetically favored, and it can promote the appearance of 
orientationally modulated phase.
Effectively it means that the softened elastic 
constant may become negative.
An important observation made by Meyer many years ago \cite{ref23} was that this softening 
can be attributed to the entropy of packing of molecules with specific shape of nonzero
steric dipole. In fact, the dipolar asymmetry of these molecules is expected to induce 
a flexopolarization effect, which couples to splay and bend director 
deformations \cite{Meyer1969} and effectively reduces the $K_{11}$ and/or $K_{33}$ 
elastic constants \cite{jakli2018physics}. For instance, 
molecules with a bow(banana) shape can reduce the $K_{33}$ bend elastic constant, 
leading to the formation of 
twist-bent and splay-bent phases \cite{Dozov_2001}.
Greco and Ferrarini were the first to explicitly demonstrate this phenomenon 
through molecular dynamic simulations 
on a system of bow-shaped molecules that interact solely
through steric interactions \cite{GrecoFerrariniPRL}.

Further studies on bow-shaped molecules,  
focusing on steric interactions, have provided successful explanations
not only for the formation 
of the $\text{N}_\text{TB}$ phase \cite{Chiappini2021,Kubala2022} but have also
demonstrated the potential for stabilizing various intermediate polar
states between $\text{N}_\text{TB}$ and $\text{N}_\text{SB}$ \cite{Chiappini2021}.

It is well-known that even slight differences in the geometry of molecules
can have a significant impact on the resulting liquid-crystalline 
self-assembly.
For example, a smectic A phase is observed in a system consisting of spherocylinders 
\cite{Veerman1990}, whereas ellipsoids with a very similar shape do not exhibit 
this behavior \cite{Frenkel1984}. This principle also applies to the recently discovered
$\text{N}_\text{F}$  and $\text{N}_\text{S}$ phases.
These phases rely on both wedge-like  molecular anisotropy 
and a strong, nearly longitudinal total molecular dipole moment 
as crucial molecular characteristics 
contributing to their stability. Specifically, the wedge-like molecular shapes
can result in a negative  $K_{11}$ splay constant  \cite{Gregorio2016},
thereby contributing to the formation of the polar splay nematic and related 
smectic phases \cite{Sebastian2022, Chen2021}.

 Numerical simulations of systems composed of vedge-like (\textit{e.g.}, pear-shaped, 
tapered, \textit{etc.}) molecules began 
in the late 1990s \cite{Stelzer1999, Stelzer2000}. In these simulations, 
shape polarity at the molecular level was induced by a soft interaction potential 
between molecules, which combined two rigidly connected centers: 
an ellipsoidal Gay-Berne potential\cite{Gay1981} 
and a spherical Lennard-Jones potential. The reported liquid crystalline 
phases in these simulations were only ordinary uniaxial nematic and smectic phases, without 
any macroscopic polarization, as they exhibited a preference for antiparallel 
local arrangement. 

Berardi, Ricci, and Zannoni developed a generalized single-site Gay-Berne potential 
to model the attractive and repulsive interactions between elongated tapered 
molecules \cite{berardi2001}. 
Through parameter adjustments and Monte Carlo simulations, they observed stable $\text{N}_\text{F}$ 
and ferroelectric smectic liquid crystals. However, while the introduction of a 
weak axial dipole did not qualitatively impact these observations, 
an increase in dipole strength resulted in the destruction of long-range 
ferroelectric ordering.
It is worth noting 
that using a standard Gay-Berne potential with an axial dipole at one end of the molecule 
resulted in the formation  of a bilayer smectic phase 
\cite{Houssa2009,longa2000structures,longa2003computer}.
Similar mesophase formation was observed in simulations of   
single-site hard pears \cite{ Barmes2003}.

Purely entropic systems constructed of 
pear-shaped molecules also exhibit 
a cubic gyroid phase \cite{Ellison2006, Schonhofer2017}.
Interestingly, the stability of this phase is highly 
sensitive to the details of the hard-core interaction. 
Specifically, the cubic gyroid phase is observed when describing 
the pear shape using two Bézier curves with the hard pear Gaussian 
overlap model (PHGO). However, it vanishes when the hard pears of 
revolution (HPR) model is used. In the PHGO model, a bilayer 
smectic phase is also observed, whereas the HPR model exhibits 
isotropic and nematic phases \cite{Schonhofer2020}.

\begin{figure}[ht]
    \centering
    \includegraphics[width=0.85\linewidth]{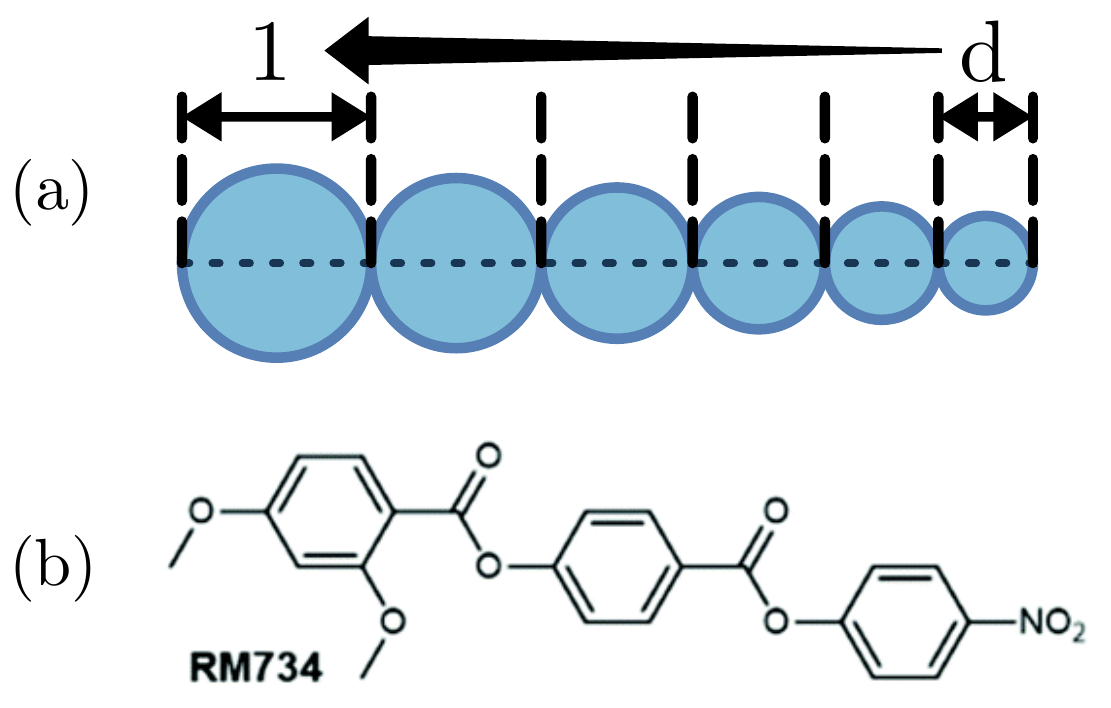}
    \caption{(a) Family of wedges used in the study; they are built of six co-linear tangent spheres with diameters increasing linearly from $d$ to 1 ($d \le 1$ is a parameter). (b) Wedge-shaped RM734 molecule known to form a polar nematic phase. Reprinted from Ref.~\cite{Mandle2022} with the permission of RSC.}
    \label{fig:shapes}
\end{figure}

In this paper, we seek to explore the fundamental 
mechanisms that can lead to the emergence of 
long-range splay and potentially polar order in liquid crystal systems. 
Specifically, we focus on the interactions among hard wedge-shaped molecules
in the absence of dipolar electrostatic or dispersion forces. 
By investigating whether such long-range order can be stabilized 
solely through entropic interactions, we aim to advance our understanding 
of the essential features of molecular interactions 
responsible for stabilizing $\text{N}_\text{F}$  and $\text{N}_\text{S}$ phases. 

To achieve this goal, we utilize a model consisting of six co-linear tangent spheres, 
resulting in a molecule with $C_{\infty v}$ (cone) symmetry [Fig.~\ref{fig:shapes}(a)].
The diameters of the spheres follow an arithmetic sequence, starting from d and 
progressing to 1. Specifically, the diameters are given by 
$d$, $(4d+1)/5$, $(3d+2)/5$, $(2d+3)/5$, $(d+4)/5$ and 1. 
Here, the parameter $d$ represents the diameter of the smallest sphere 
and serves as a descriptor for the shape of the molecule. 
When $d$ equals $1$, the molecule reduces to the linear 
tangent hard-sphere model (LTHS) as described by Vega \textit{et al.} in Ref.~\cite{Vega2001}.

We have chosen this model to capture essential features of the effective shape exhibited 
by the RM734 molecule 
(4-[(4-nitrophe-noxy)carbonyl]phenyl2,4-dimethoxybenzoate) [Fig.~\ref{fig:shapes}(b)].
As previously mentioned, the RM734 mesogen has been shown to stabilize
$\text{N}_\text{F}$ and $\text{N}_\text{S}$ phases \cite{Mandle2021, Mandle2022}. 
By utilizing a model with similar shape characteristics, we aim to gain 
further insight into the relationship between entropy of packing and the resulting 
self-organization in liquid crystal systems. To this end, we 
investigate the phase diagram and properties of stable structures 
using Monte Carlo integration. Additionally, we 
compare our results with those obtained from previous studies on 
soft- and hard-core pear models. Furthermore, we examine 
how the gradient of a molecule's diameter 
influences the presence and extent of the observed phases. 

Finally, we recognize that the anisotropic polar 
shape of the molecules may give rise to non-trivial 
dense configurations. We are particularly interested 
in exploring whether these configurations exhibit 
crystal-like or glass-like structures, considering the potential 
competition between different types of lattices \cite{Chen2014, Klotsa2018, Teich2019}. 

The remainder of this paper is organized as
follows: In Sec. II A we provide a brief description of the Monte Carlo integrator
used. In
Sections II B and II C, we introduce the order parameters and correlation functions
used to monitor properties of equilibrium structures. The results obtained from Monte Carlo 
simulations are presented in Sec. III. 
In Sec. IV, we formulate the Parsons-Lee Density Functional Theory to study polar
ordering in the liquid-crystalline regime. We analyze some equilibrium and metastable 
phases in detail.  Lastly, we provide a discussion and outlook in
Sec. V.  Appendix A contains remarks about non-tilted hexagonal configurations of close-packed linear tangent  hexamers.

\section{Methods}

\subsection{Monte Carlo simulations} \label{sec:MC}

We assumed hard-core interactions between molecules. The equilibrium phases were classified as a function of the shape parameter $d$ and the packing fraction $\eta$. The latter one is a natural choice for purely steric repulsion. System snapshots were obtained numerically using the Monte Carlo scheme \cite{Allen2017} implemented in our RAMPACK software package (see Sect. \textit{Code availability}). Integration was carried out in the $NpT$ ensemble. For hard-core interactions, only a ratio $p/T$ of pressure and temperature is an independent parameter and can be used to control the packing fraction $\eta$. To allow relaxation of the full viscous stress tensor (including the sheer part), we used a triclinic simulation box with periodic boundary condition. For a system of $N$ molecules, a full MC cycle consisted of $N$ rototranslation moves, $N/10$ flip moves, and a single box move. In a rototranslation move, a single shape was chosen at random, translated by a random vector, and rotated around the random axis by a random angle (clockwise and anticlockwise rotations were equally probable to preserve the detailed balance condition). If the move introduced an overlap, it was always rejected and accepted otherwise. Flip moves were performed in a similar way to rototranslation moves; however, instead of random translation and rotation, the molecule was rotated by $180^\circ$ around its geometric center. This type of move facilitated easier sampling of the phase space of the system, especially for high $\eta$. For a box move, the three vectors $\vb{b}_1$, $\vb{b}_2$, $\vb{b}_3$ that span the box were perturbed by small random vectors. The move was rejected if any overlaps were introduced. Otherwise, it was accepted according to the Metropolis-Wood criterion with probability \cite{Wood1968,Wood1968jcp}
\begin{equation} \label{eq:metropolis}
    \min \qty{1, \exp\qty(N \log\frac{V}{V_0}  - \frac{p\Delta V}{T})},
\end{equation}
where $\Delta V = (V - V_0)$ and $V_0$, $V$ are, respectively, the volume of the box before and after the move. The perturbation ranges were adjusted during the thermalization phase to achieve an acceptance probability of around 0.15. To accelerate simulations in a modern multi-threaded environment, we used domain decomposition technique \cite{Uhlherr2003} for molecule moves and we parallelized independent overlap checks for volume moves.

To scan the full phase sequence, from isotropic liquid to crystal, we used $p/T$ ratios corresponding to packing fraction covering $\eta \in [0.3, 0.58]$ for $d \in [0.4, 1.0]$. First, to roughly determine the phase boundaries, preliminary simulations were performed by gradually compressing a small system of $N = 400$ molecules in a cubic box from a highly diluted simple cubic lattice. For each $p/T$, the integration consisted of the thermalization run with $9.5 \times 10^6$ full MC cycles and the production run with $0.5 \times 10^6$ cycles to gather averages. The final snapshot of a run was used as a starting point for the next with a slightly higher $p/T$. Using the results as guidance, the main simulations were performed on a much larger system with $N > 5000$ in a triclinic box. The initial configuration in the whole range of $d$ was smectic A with $\eta \approx 0.45$ (see Sec.~\ref{sec:results} for the description of phases) prepared by thermalizing different types of slightly diluted crystals for (1-5)$\times 10^8$ cycles. Initial configurations were then independently compressed or expanded to all target densities in parallel. Thermalization runs were performed for (0.9-4.5)$\times 10^8$ cycles, while production runs were performed for (0.1-0.5)$\times 10^8$ cycles. Additionally, in order to estimate maximal packing fractions, the densest configurations for each $d$ were compressed under exponentially increasing pressure for $3 \times 10^8$ cycles, reaching $p/T=10^4$ at the end.

\subsection{Order parameters}

Phases in the system can be easily classified using a carefully chosen set of order parameters, whose values have jumps on the boundaries of phase transitions. The nematic order along the director $\vu{n}$ is detected by the average value $\expval{P_2(\vu{a} \vdot \vu{n})}$ of the second-order Legendre polynomial, where $\vu{a}$ is the long axis of the molecule. Director $\vu{n}$ can be inferred directly from the system using the second-rank $\vb{Q}$ tensor \cite{Eppenga1984}, which can be numerically computed as
\begin{equation} \label{eq:Q}
    \vb{Q} = \frac{1}{N} \sum_{i=1}^{N} \frac{3}{2}\qty(\vu{a}_i \otimes \vu{a}_i - \frac{1}{3}),
\end{equation}
where the summation is done over all molecules in a single snapshot. $P_2$ is then the eigenvalue of $\vb{Q}$ with the highest magnitude and $\vu{n}$ -- the corresponding eigenvector. Ensemble averaged $\expval{P_2}$ is calculated by averaging $P_2$ over non-correlated system snapshots. The nematic order parameter has a minimal value $-0.5$, when all molecules are perpendicular to $\vu{n}$, and reaches its maximum 1 for molecules perfectly aligned with $\vu{n}$ (please note that $\vu{n}$ and $-\vu{n}$ directions are equivalent). In a disordered system $\expval{P_2} = 0$.

Density modulation can be quantitatively described by the smectic order parameter $\expval{\tau}$ \cite{deGennesBook}. It is defined as
\begin{equation} \label{eq:tau}
    \expval{\tau} = \frac{1}{N} \expval{\abs{\sum_{i=1}^{N} \exp(i \vb{k} \vdot \vb{r}_i)}},
\end{equation}
where $\vb{k}$ is the modulation wavevector compatible with PBC and $\vb{r}_i$ is the center of the $i^\text{th}$ molecule. As the drift of the whole system is a Goldstone mode, the absolute value $\abs{\cdots}$ is taken before the ensemble averaging to eliminate it. All possible $\vb{k}$ can be enumerated using reciprocal box vectors $\vb{g}_1, \vb{g}_2, \vb{g}_3$ \footnote{$\vb{g}_i$ can be read off as rows of matrix $G=2\pi\vb{M}^{-1}$, where columns of $\vb{M}$ are vectors spanning simulation box} and taking linear combinations of them with integer coefficients $h, k, l$ (Miller indices \cite{de2016basic}): $\vb{k} = h\vb{g}_1 + k\vb{g}_2 + l\vb{g}_3$. Here, as the initial configuration is always a smectic with six layers stacked along the $z$ axis, $hkl=006$. The smectic order ranges from 0 for a homogeneous system to 1 for a perfectly layered one.

Another feature of the system that is measured in the study is the hexatic order appearing for high packing fractions $\eta$, where molecules tend to form hcp-like structures. The local hexatic order can be measured using the so-called hexatic bond order parameter $\expval{\psi_6}$ \cite{Nelson2012}. For a two-dimensional system, it is defined as
\begin{equation} \label{eq:psi}
    \expval{\psi_6} = \frac{1}{N} \expval{\sum_{i=1}^{N} \frac{1}{6} \abs{ \sum_{j=1}^{6} \exp(6i\phi_{ij})}},
\end{equation}
where $\phi_{ij}$ is the angle between an arbitrary axis in the plane and the vector that joins the center of the $i^\text{th}$ molecule with its $j^\text{th}$ nearest neighbor. It can be generalized to three-dimensional systems by projecting the positions of molecules onto the nearest smectic layers and computing $\psi_6$ within the planes defined by them. Random points give $\expval{\psi_6} \approx 0.37$, while a perfect hexatic order yields $\expval{\psi_6} = 1$. The local hexatic order can also be computed for a system without layers by projecting all centers on a single plane.


\subsection{Correlation functions}

Additional insight into both global properties and supramolecular structures is given by correlation functions. The first is a standard radial distribution function \cite{Stone1978}, which can be defined in a computationally friendly way as
\begin{equation} \label{eq:rdf}
    \rho(r) = \expval{\expval{\frac{\dd{N}(r, r+\dd{r})}{4 \pi r^2 \dd{r} \cdot (N/V) }}_N},
\end{equation}
where $\dd{N}(r, r+\dd{r})$ is the number of molecules whose distance from a selected single molecule lies in the range $(r, r+\dd{r})$, $\dd{r}$ is the numerical size of the bin, $N$ is the total number of molecules and $V$ is the volume of the system. It is then averaged over all molecules $\expval{\cdots}_N$ and over independent snapshots $\expval{\dots}$. It is normalized in such a way that, for a disordered isotropic system, it approaches 1 for $r \to \infty$. In systems with long-range translational order, $\rho(r)$ has a series of numerous minima and maxima.

In a layered system, one can also measure the layer-wise radial distribution function in the direction orthogonal to $\vb{k}$
\begin{equation} \label{eq:rdf_L}
    \rho_\perp(r_\perp) = \frac{1}{n_L} \expval{\sum_{i=1}^{n_L} \expval{\frac{\dd{N}_i(r_\perp, r_\perp+\dd{r_\perp})}{2 \pi r_\perp \dd{r_\perp} \cdot (N/S) }}_{N_i}},
\end{equation}
where $n_L$ is a number of layers, $\dd{N}_i(r_\perp, r+\dd{r_\perp})$ is the number of molecules in $i^\text{th}$ layer whose distance from a selected single molecule calculated along layer's plane lies in the range $(r_\perp, r_\perp+\dd{r_\perp})$, and $S$ is total surface area of all layers \footnote{Please note that for general Miller indices $hkl$ some layers may be connected through PBC. $n_L$ is a number of disjoint layers and is equal to the highest common divisor of $h, k, l$, while $S = V \norm{h\vb{g}_1 + k\vb{g}_2 + l\vb{g}_3}$.}. In the end, it is averaged over all molecules in the layer $\expval{\cdots}_{N_i}$, all layers $(1/n_L)\sum_{i=1}^{n_L}\cdots$ and uncorrelated system snapshots $\expval{\cdots}$.

As the results will show, the system develops a nontrivial polar metastructure. To quantify it, we use the layer-wise radial polarization correlation \cite{Stone1978}, defined alike $\rho_\perp(r_\perp)$:
\begin{equation} \label{eq:S110}
    S^{110}_\perp(r_\perp) = \frac{1}{n_L} \expval{\sum_{i=1}^{n_L} \expval{\vu{a}_{i_j} \cdot \vu{a}_{i_k}}_{i_j i_k}},
\end{equation}
where $\expval{\cdots}_{i_j i_k}$ denotes the average over all molecules in the $i^\text{th}$ layer, whose centers' distance along this layer lies in the $(r_\perp, r+\dd{r_\perp})$ range.

The range of splay correlations can be quantified by the conditional probability $P(\theta|r_\perp)$ of finding two molecules with angle $\theta$ between their molecular axes $\vu{a}_i$ at a transversal distance $r_\perp$, normalized as
\begin{equation}
    \int_{0^\circ}^{90^\circ} P(\theta|r_\perp) \dd{\theta} = 1, \qquad \forall r_\perp.
\end{equation}
To be consistent with the polarization correlation function, this quantity will also be calculated layer-wise. As splay corresponds to the radial spread of director field lines, the most probable angle $\theta$ should grow with $r_\perp$ in systems with non-zero splay deformation mode.

\section{Results} \label{sec:results}

\begin{figure}[ht]
    \centering
    \includegraphics[width=0.8\linewidth]{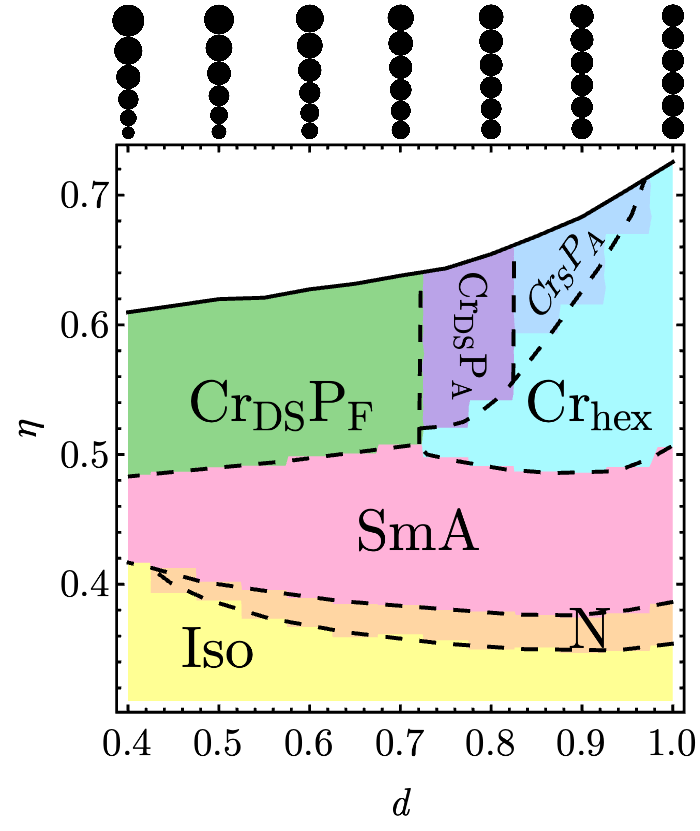}
    \caption{Phase diagram of the system in $(d, \eta)$ space. The following liquid phases were recognized: isotropic (Iso), nematic (N) and smectic A (SmA) as well as crystalline ones: hexagonal crystal ($\text{Cr}_\text{hex}$), antiferroelectric double splay crystal ($\text{Cr}_\text{DS}\text{P}_\text{A}$), antiferroelectric splay crystal ($\text{Cr}_\text{S}\text{P}_\text{A}$) and ferroelectric double splay crystal ($\text{Cr}_\text{DS}\text{P}_\text{F}$). Black solid line at the top boundary of the phase diagram represents the maximal packing fraction $\eta$ for a given $d$ value.}
    \label{fig:pd}
\end{figure}

\begin{figure}[ht]
    \centering
    \includegraphics[width=0.8\linewidth]{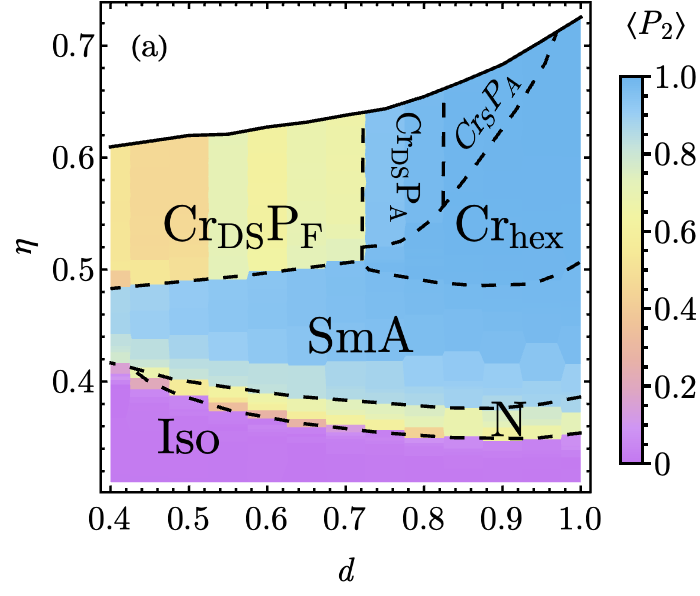}
    \includegraphics[width=0.8\linewidth]{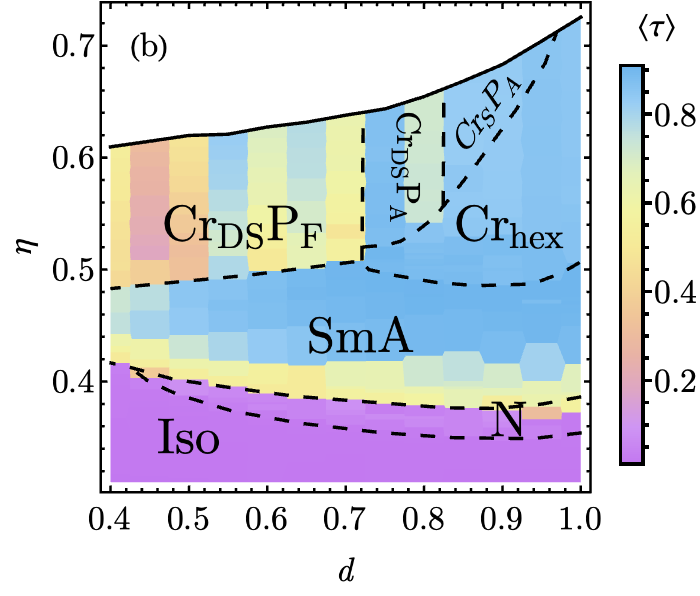}
    \includegraphics[width=0.8\linewidth]{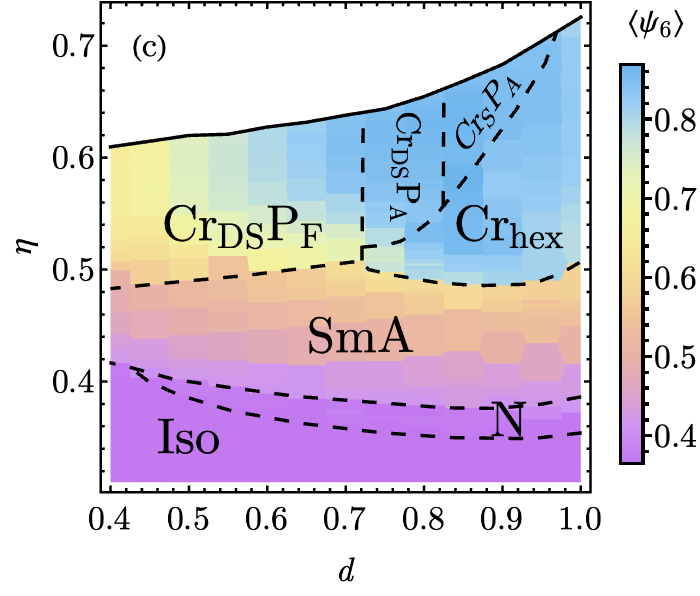}
    \caption{Ensemble averages of observables as a function of the smallest sphere's diameter $d$ and packing fraction $\eta$. (a) Nematic order $\expval{P_2}$, (b) smectic order $\expval{\tau}$, (c) local hexatic order $\expval{\psi_6}$.}
    \label{fig:pd_obs}
\end{figure}
    
\begin{figure}[htb]
    \centering
    \includegraphics[width=0.95\linewidth]{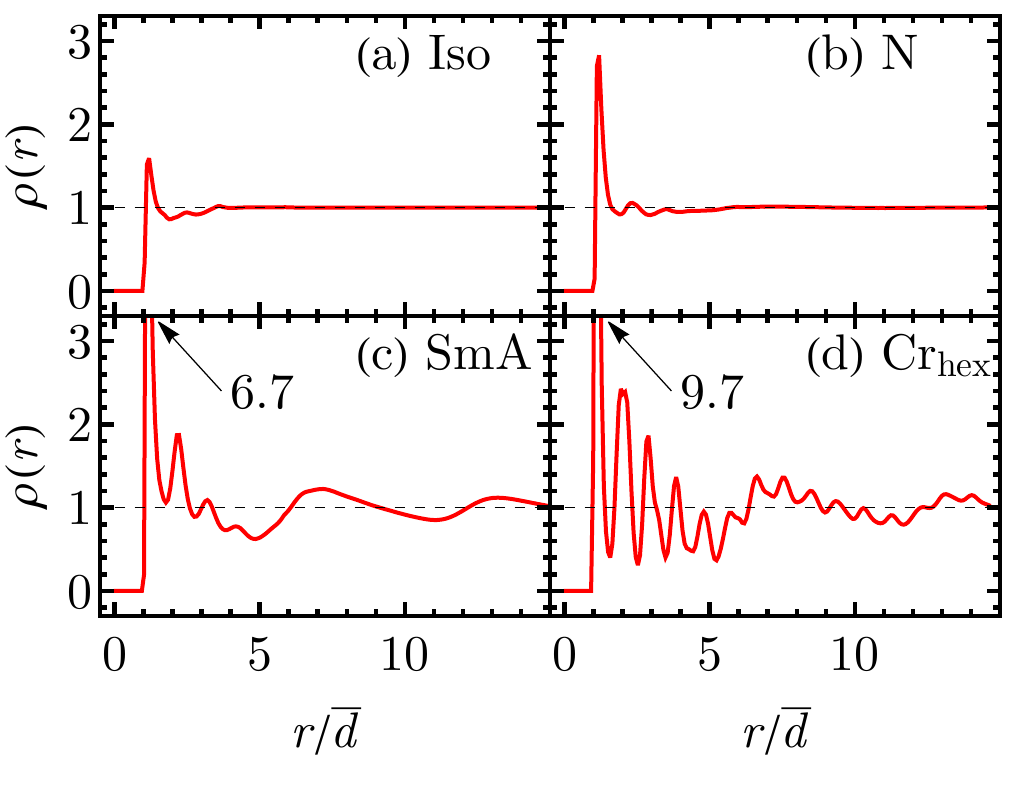}
    \caption{Radial distribution function $\rho(r)$ for (a) Iso [$(d, \eta) = (0.6, 0.34)$], (b) N [$(d, \eta) = (0.5, 0.39)$], (c) SmA [$(d, \eta) = (0.5, 0.46)$], (d) $\text{Cr}_\text{hex}$ [$(d, \eta) = (0.95, 0.51)$] phases. Distance $r$ is scaled by the average diameter of balls in the molecule $\bar{d} = (d+1)/2$. Correlation peaks are clipped on panels (c) and (d), however their values are shown inside the plots.}
    \label{fig:rho}
\end{figure}
    
\begin{figure}[htb]
    \centering
    \includegraphics[width=0.95\linewidth]{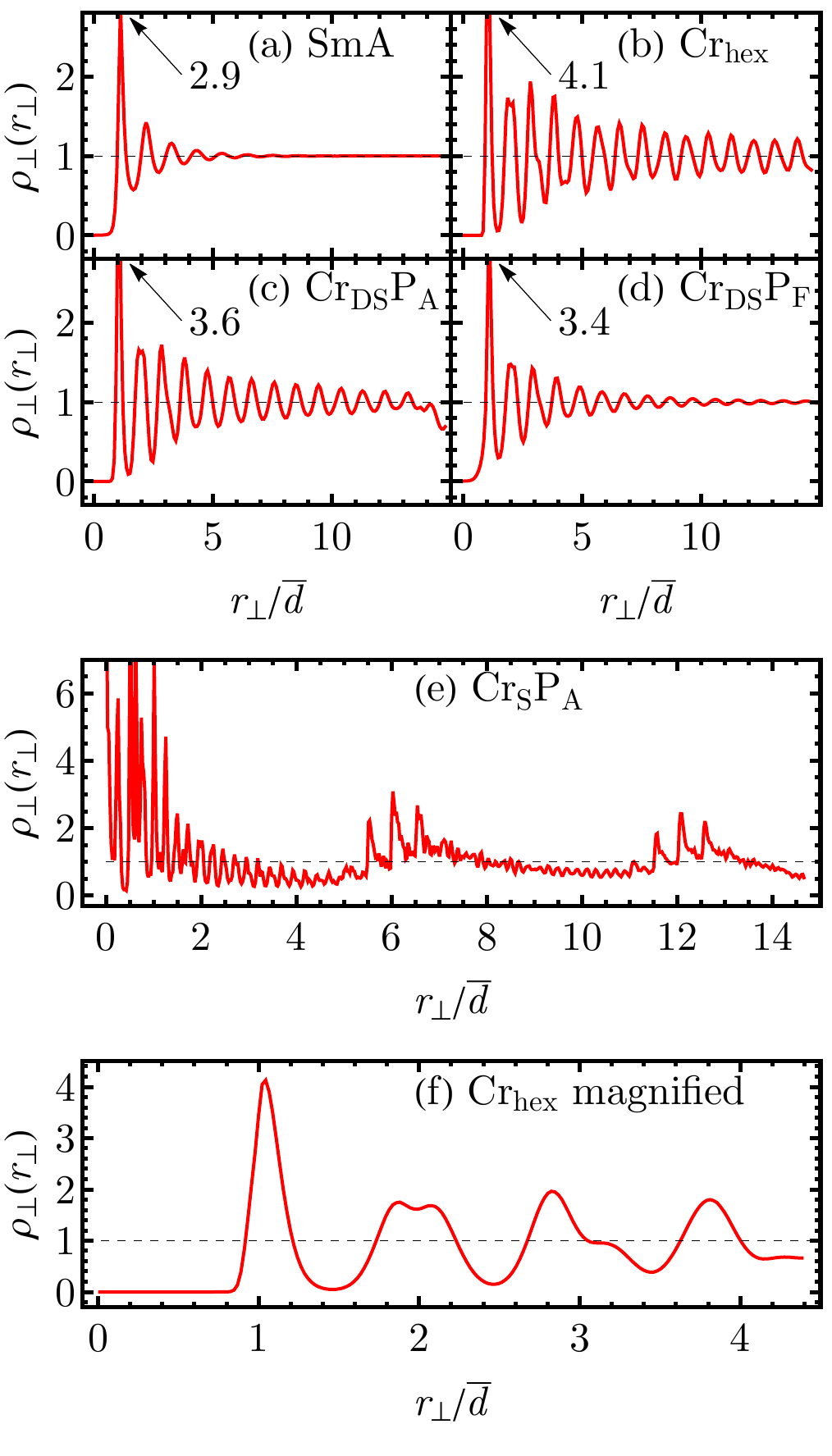}
    \caption{Layer-wise distribution function $\rho_\perp(r_\perp)$ for (a) SmA [$(d, \eta) = (0.5, 0.46)$], (b) $\text{Cr}_\text{hex}$ [$(d, \eta) = (0.95, 0.51)$], (c) $\text{Cr}_\text{DS}\text{P}_\text{A}$ [$(d, \eta) = (0.75, 0.52)$], (d) $\text{Cr}_\text{DS}\text{P}_\text{F}$ [$(d, \eta) = (0.6, 0.51)$] and (e) $\text{Cr}_\text{S}\text{P}_\text{A}$ [$(d, \eta) = (0.9, 0.68)$] phases. Moreover, panel (f) shows magnification of a part of panel (b). Distance $r_\perp$ is scaled by the average diameter of balls in the molecule $\bar{d} = (d+1)/2$. Correlation peaks are clipped on panels (b), (c) and (d), however their values are shown inside the plots.}
    \label{fig:rhoL}
\end{figure}
    
\begin{figure}[htb]
    \centering
    \includegraphics[width=0.95\linewidth]{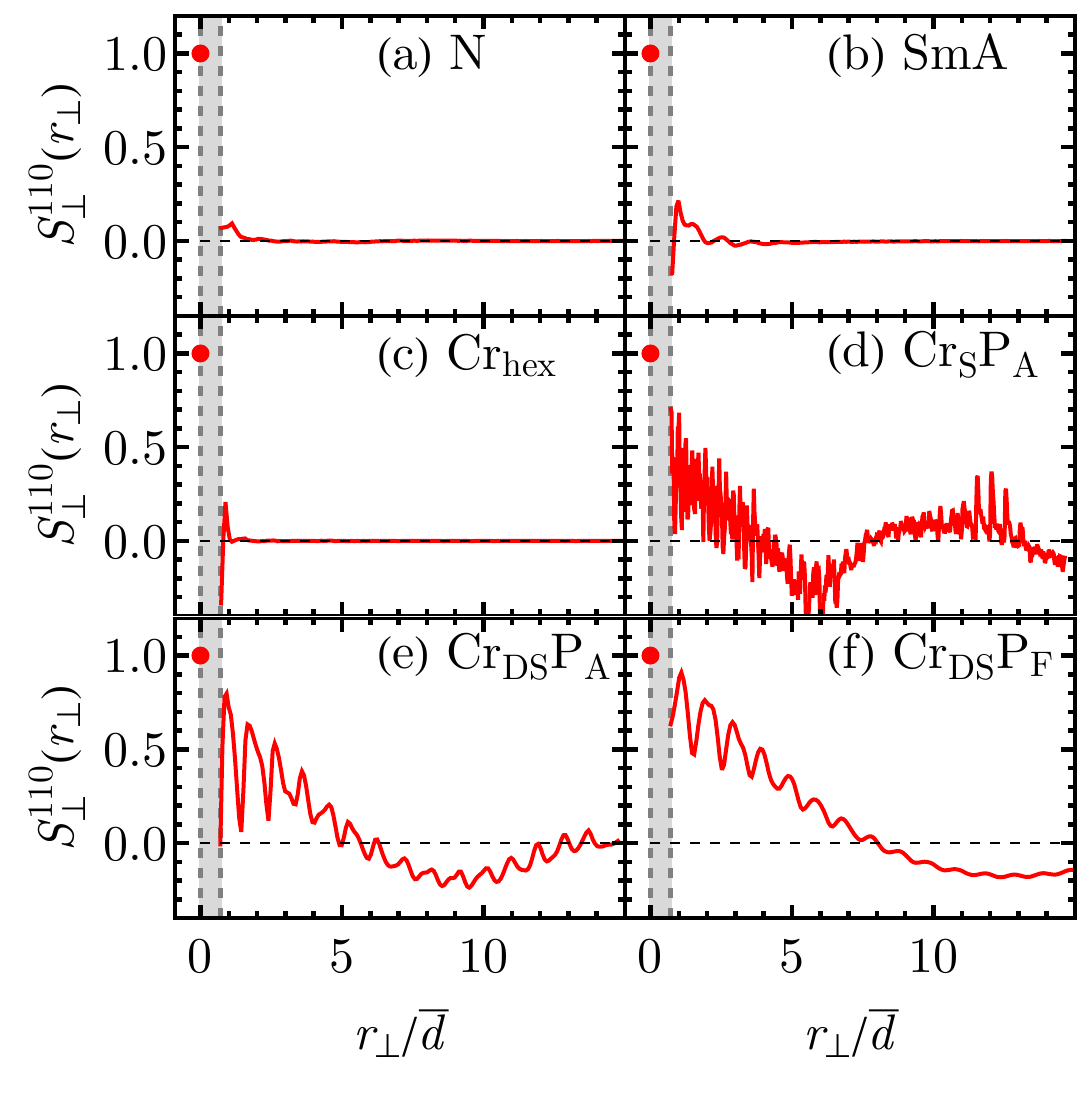}
    \caption{Layer-wise correlation function $S^{110}_\perp(r_\perp)$ for (a) N [$(d, \eta) = (0.5, 0.34)$], (b) SmA [$(d, \eta) = (0.5, 0.46)$], (c) $\text{Cr}_\text{hex}$ [$(d, \eta) = (0.95, 0.51)$], (d) $\text{Cr}_\text{S}\text{P}_\text{A}$ [$(d, \eta) = (0.9, 0.68)$], (e) $\text{Cr}_\text{DS}\text{P}_\text{A}$ [$(d, \eta) = (0.75, 0.52)$], (f) $\text{Cr}_\text{DS}\text{P}_\text{F}$ [$(d, \eta) = (0.6, 0.51)$] phases. Distance $r_\perp$ is scaled by the average diameter of balls in the molecule $\bar{d} = (d+1)/2$. Due to excluded volume, there are scarcely any molecules in range $r_\perp \in [0, \bar{d}]$, therefore this area is grayed out.}
    \label{fig:S110}
\end{figure}
    
\begin{figure}[htb]
    \centering
    \includegraphics[width=0.75\linewidth]{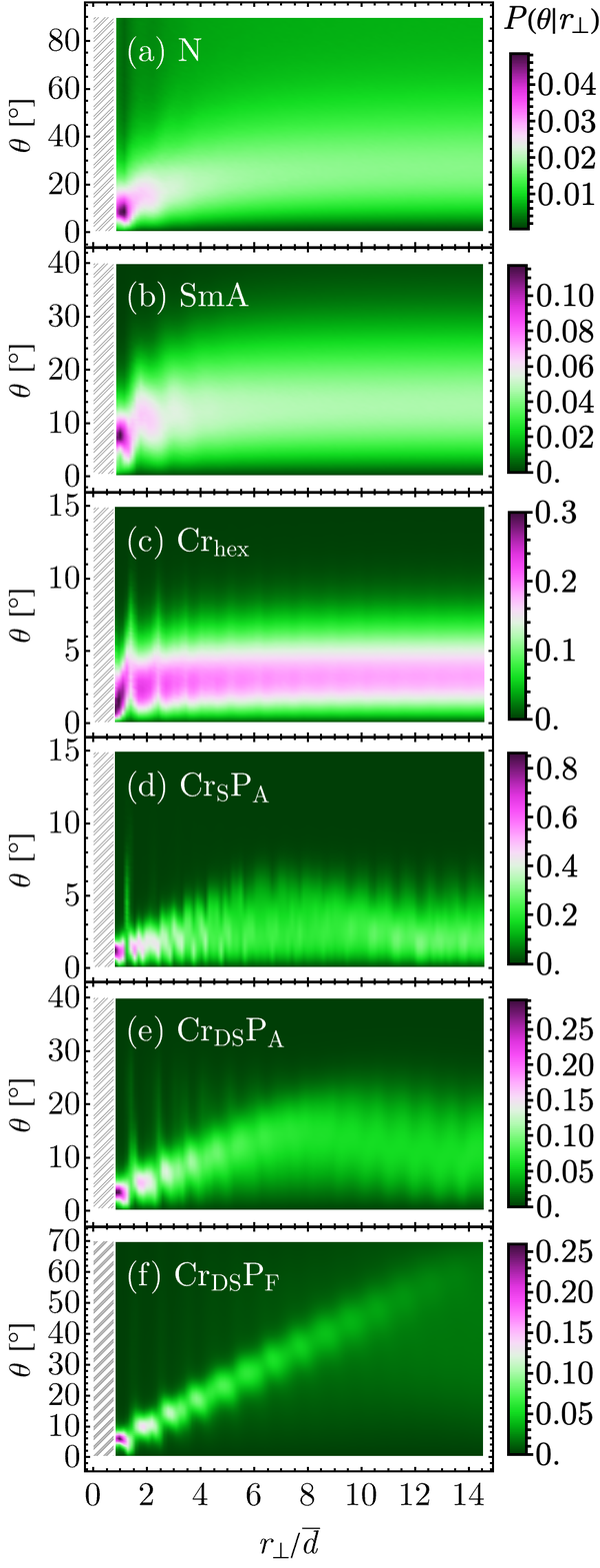}
    \caption{The conditional probability $P(\theta|r_\perp)$ for (a) N [$(d, \eta) = (0.5, 0.34)$], (b) SmA [$(d, \eta) = (0.5, 0.46)$], (c) $\text{Cr}_\text{hex}$ [$(d, \eta) = (0.95, 0.51)$], (d) $\text{Cr}_\text{S}\text{P}_\text{A}$ [$(d, \eta) = (0.9, 0.68)$] (e) $\text{Cr}_\text{DS}\text{P}_\text{A}$ [$(d, \eta) = (0.75, 0.52)$] and (f) $\text{Cr}_\text{DS}\text{P}_\text{F}$ [$(d, \eta) = (0.6, 0.51)$] phases.}
    \label{fig:angle}
\end{figure}

\begin{figure*}[htbp]
    \centering
    \includegraphics[width=0.8\linewidth]{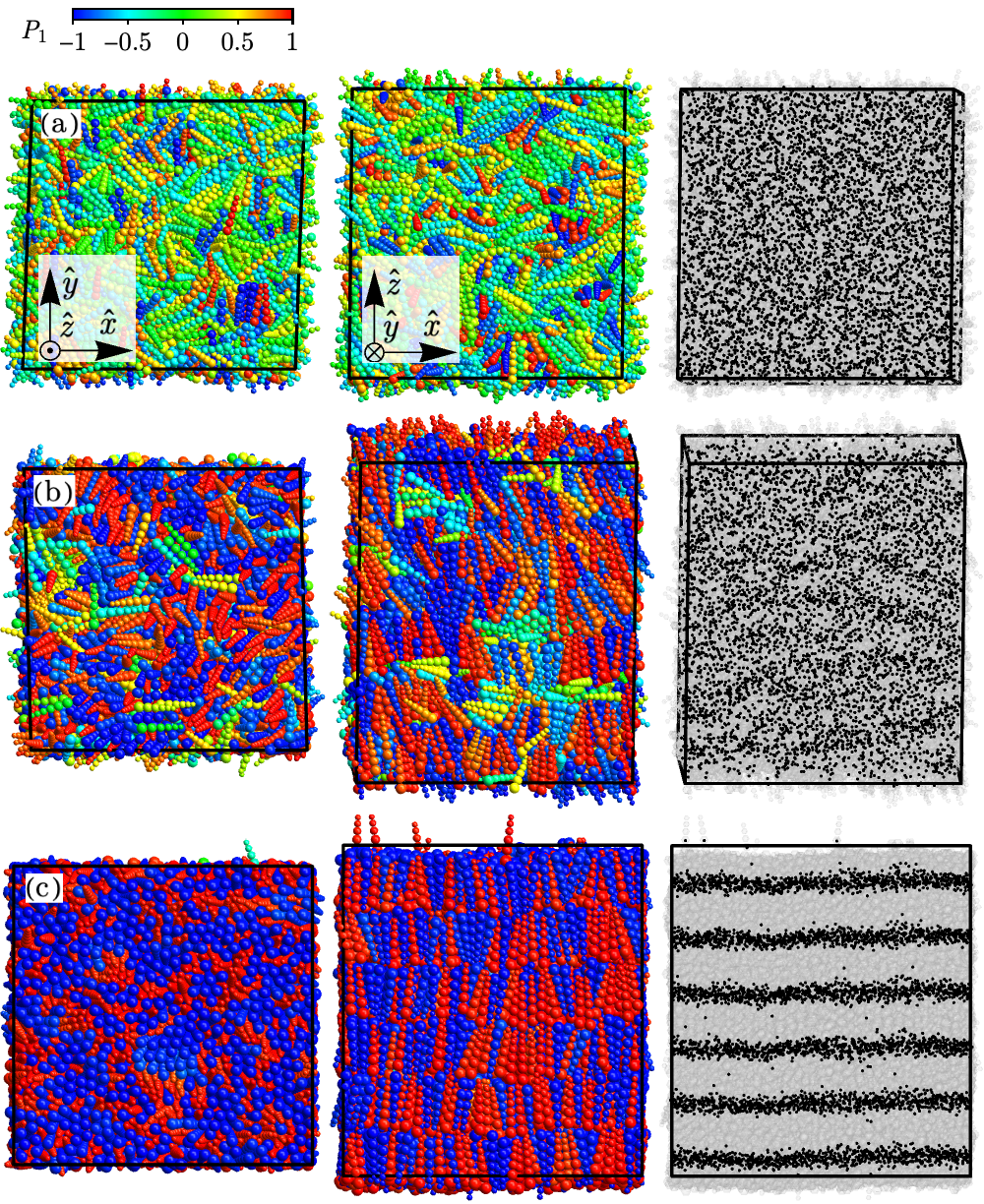}
    \caption{System snapshots of all liquid phases in the system. Rows correspond to, (a) Iso [$(d, \eta) = (0.6, 0.34)$], (b) N [$(d, \eta) = (0.5, 0.39)$] and (c) SmA [$(d, \eta) = (0.5, 0.46)$] phases. First column presents the top view of the snapshot ($xy$ plane), while the second one the side view ($xz$) plane in a way that bottom edge of the top view corresponds to the top edge of the side view. Molecules are color-coded according to orientation of their long axis $\vu{a}_i$ with respect to the global director $\vu{n}$ as per first order Legendre polynomial $P_1(\vu{a}_i \cdot \vu{n}) = \vu{a}_i \cdot \vu{n}$, allowing one to discern opposite polarizations. In the third column molecules' centers are marked as black dots, and the simulation box is oriented in the same way as the second column (side view).}
    \label{fig:pack_liquid}
\end{figure*}

\begin{figure*}[htbp]
    \centering
    \includegraphics[width=0.65\linewidth]{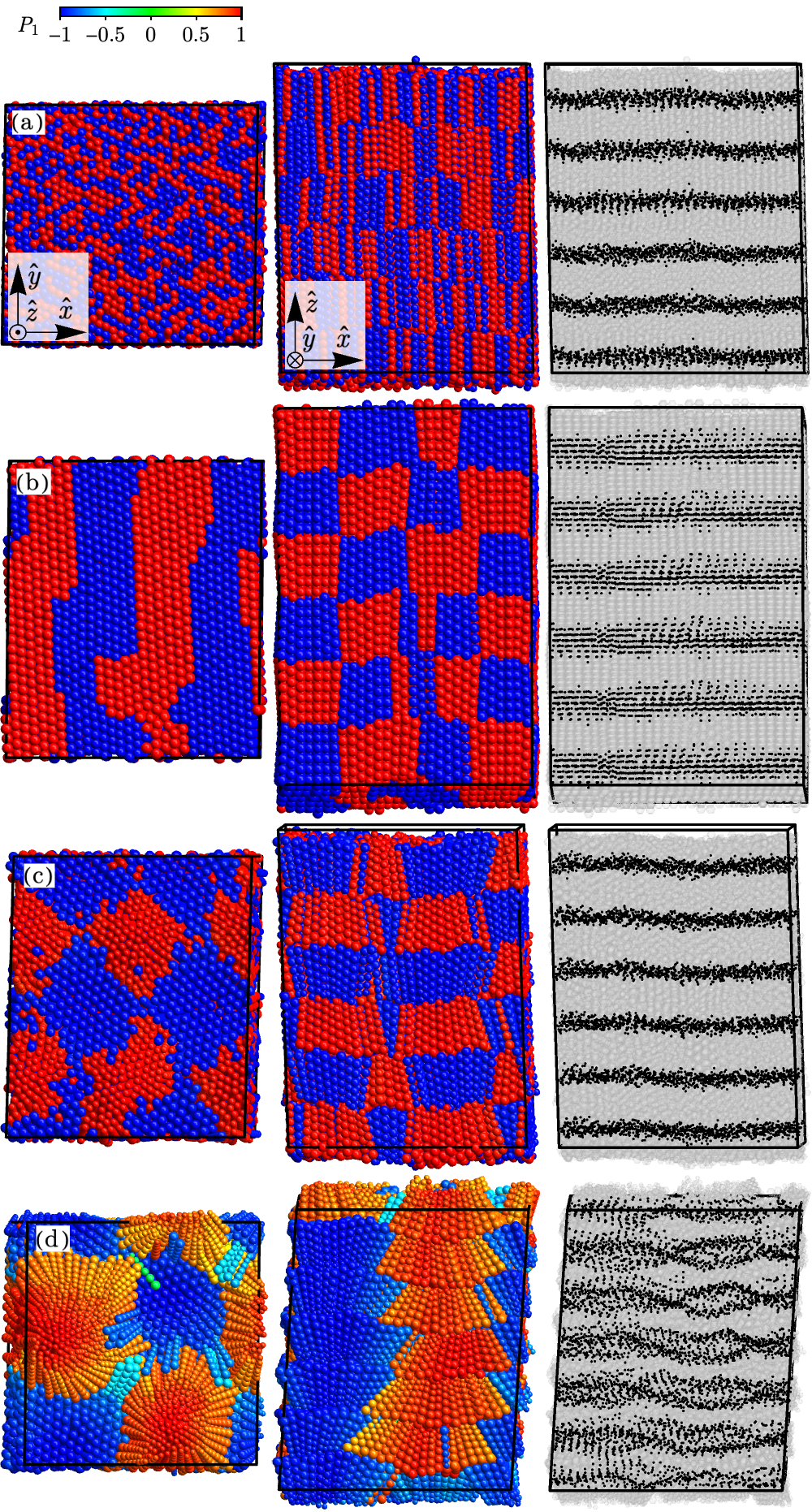}
    \caption{System snapshots of all crystalline phases in the system. Rows correspond to, respectively, (a) $\text{Cr}_\text{hcp}$ [$(d, \eta) = (0.95, 0.51)$], (b) $\text{Cr}_\text{S}\text{P}_\text{A}$ [$(d, \eta) = (0.9, 0.68)$], (c) $\text{Cr}_\text{DS}\text{P}_\text{A}$ [$(d, \eta) = (0.75, 0.52)$] and (d) $\text{Cr}_\text{DS}\text{P}_\text{F}$ [$(d, \eta) = (0.6, 0.51)$] phases. Columns' order and color coding are the same as in Fig.~\ref{fig:pack_liquid} -- bluefrom the left: top view, side view, side view of mass centers. The average number of molecules in clusters is as follows: (b) $\approx 7$ per stripe per $XZ$ section, (c) $\approx 110$ per cluster, (d) $\approx 220$ per column per layer.}
    \label{fig:pack_crystal}
\end{figure*}

Using the method described in Section \ref{sec:MC}, we were able to recognize all phases for $d \in [0.4, 1.0]$ and $\eta \in [0.3, 0.58]$. There are three liquid phases: isotropic liquid (Iso), nematic (N), smectic A (SmA) and four crystalline phases: hexagonal crystal ($\text{Cr}_\text{hex}$), antiferroelectric splay crystal ($\text{Cr}_\text{S}\text{P}_\text{A}$), antiferroelectric double splay crystal ($\text{Cr}_\text{DS}\text{P}_\text{A}$) and ferroelectric double splay crystal ($\text{Cr}_\text{DS}\text{P}_\text{F}$). The phase diagram is presented in Fig.~\ref{fig:pd}, the order parameters are shown in Fig.~\ref{fig:pd_obs}, while Figs.~\ref{fig:rho},\ref{fig:rhoL},\ref{fig:S110} contain correlation functions. Moreover, representative equilibrium snapshots of all phases can be seen in Figs.~\ref{fig:pack_liquid},\ref{fig:pack_crystal}. The phases for all sampled pairs $(d, \eta)$ were manually classified using order parameters and visual inspection of system snapshots. They are thoroughly analyzed in the following sections.
\subsection{Liquid phases}
For the lowest packing densities $\eta$,  the system forms an isotropic liquid phase without any
long-range translational or orientational ordering. An example snapshot of this phase 
is shown in Fig.~\ref{fig:pack_liquid}(a). In this phase, the values of the nematic order parameter 
$\expval{P_2}$ and smectic order parameter $\expval{\tau}$ are close to zero, 
indicating the absence of alignment or layering.  The hexatic order parameter $\expval{\psi_6}$ 
also has a minimal value (see Fig.~\ref{fig:pd_obs}), indicating a lack of hexagonal arrangement.
The radial distribution function [Fig.~\ref{fig:rho}(a)] shows only local correlations, 
which disappear for distances $r/\bar{d} > 4$, where $\bar{d} = (1+d)/2$ is the average diameter of balls 
that build the molecule. The maxima and minima are around integer 
multiples of $\bar{d}$: $r/\bar{d} = 1, 2, \dots$, which can be attributed to the excluded volume effects.

Upon compression, in the range of $d \in [0.5, 1]$ the nematic phase appears stable [see Fig.~\ref{fig:pack_liquid}(b)]. 
In this phase, the molecules orient, on average, along a preferred direction called the director $\vu{n}$, 
while maintaining liquid-like positions of their centers of mass. The nematic order parameter $\expval{P_2}$ 
exhibits a sharp increase at the phase boundary, jumping from 0 to 0.5-0.6 [Fig.~\ref{fig:pd_obs}(a)].
The smectic order parameter $\expval{\tau}$ and the hexatic order parameter $\expval{\psi_6}$, however, 
remain nearly minimal [Fig.~\ref{fig:pd_obs}(b, c)]. The range of the nematic phase in terms of 
packing fraction ($\Delta\eta$) is approximately $\Delta\eta \approx 0.03$ for all $d$, 
but it drops to 0 for the lowest values of $d$. 
The Iso-N phase boundary reaches a minimal value of $\eta$ at around $\eta \approx 0.35$ for highest value of $d$ 
and then moves upward towards a triple point $(d, \eta) = (0.5, 0.4)$ as $d$ decreases.

It is expected that as the value of $d$ decreases, the molecules become shorter (less anisotropic).
Consequently, a higher pressure is needed to induce ordering in the system. 
As $\eta$ increases, the nematic order parameter $\expval{P_2}$ reaches values of 0.6-0.8. 
These values are relatively high compared to the typical range observed in experiments $[0.3, 0.7]$ 
\cite{Chandrasekhar1980}. However, they are comparable to values reported in computational studies of 
other hard-shaped molecules \cite{McGrother1996, Vega2001, Kubala2022}. 

The radial distribution function $\rho(r)$ [Fig.~\ref{fig:rho}(b)] exhibits a series 
of maxima and minima around $r/\bar{d} = 1, 2, \dots$, which is typical for liquids \cite{Allen2017}. 
However, compared to the Iso phase, the first maximum is more than two times higher, indicating stronger correlations.
These correlations diminish at larger distances, for $r/\bar{d} > 6$. 

In the ferroelectic nematic ($\text{N}_\text{F}$) and splay nematic ($\text{N}_\text{S}$) phases, 
a long-range polarization order is present. Thus, it is imperative to quantify it in our system. 
Although layers are not present in the nematic phase, it is still possible to compute $S_\perp^{110}(r_\perp)$. The simplest soultion would be to project all molecules onto a single plane. However, in order to facilitate capturing local corelations, we divided the box into six identical slices (consistent with the number of layers in smectic and crystalline phase) and proceeded to compute $S_\perp^{110}(r_\perp)$ as it would be done for layered structures.
The results are presented in Fig.~\ref{fig:S110}(a). It is evident that the polarization correlations are 
relatively weak and short-ranged, vanishing completely for $r_\perp > 2\bar{d}$. 
Similarly, the splay correlations [Fig.~\ref{fig:angle}(a)] also exhibit a local range. 
The most probable angle in the system is approximately $\theta \approx 25^\circ$, 
although the maximum has a broad distribution.

The Type A smectic phase (SmA) is formed over the Iso phase for $r \in [0.4, 0.5)$ and over the N phase 
for $r \in [0.5, 1.0]$. This phase is characterized by a significant jump in the smectic order parameter
$\expval{\tau}$ [Fig.~\ref{fig:pd_obs}(b)] to around 0.5,  indicating the emergence of well-defined 
layers [as seen in the snapshot from Fig.~\ref{fig:pack_liquid}(c)]. 
The director $\vu{n}$ is parallel to the smectic wavevector $\vb{k}$, indicating the preferred 
orientation of the molecules within the layers.

In the SmA phase, there is a slight increase in the value of the parameter
$\expval{\psi_6}$ compared to the isotropic and nematic phases, ranging from 0.38 to 0.45. 
In the high packing fraction regime of this phase, $\expval{\psi_6}$ reaches values of 0.6-0.65,
indicating some degree of local hexatic ordering within the layers. 
However, it is important to note that long-range bond order is not present in this phase \footnote{Long-range bond order can be quantified using the global bond order parameter, where the modulus is on the outside of the outer $\sum_{i=1}^{N}$ sum [cf. Eq.~\eqref{eq:psi}]; it is non-zero if local hexagons are in phase, which is not the case in our system (result not shown).}, 
distinguishing it from other smectic phases such as Type B smectic \cite{Birgeneau1978}. 
The nematic order parameter $\expval{P_2}$ in SmA 
increases with increasing $\eta$, approaching a value close to 1. However, there is no sudden jump in
$\expval{P_2}$ at the N-SmA boundary. 
Similar to the N phase, the boundary moves upward with decreasing $d$, which can again be 
explained by a lower anisotropy of the molecules. 

The radial distribution function $\rho(r)$ [see Fig.~\ref{fig:rho}(c)] in SmA
exhibits a more complex structure compared to the Iso and N phases. 
It shows two superimposed sequences of minima and maxima. 
The first sequence, covering the entire range of $r/\bar{d}$, 
exhibits maxima at $r/\bar{d} \approx 7$ and  14, 
corresponding to the distances between the layers. The second sequence has a spacing of approximately $\bar{d}$ 
and vanishes for $r/\bar{d} > 6$. The first maximum in the second sequence is sharp and has a value of
$\rho(r) = 6.7$, corresponding to short-range ordering of molecules within the layers. 

The layer-wise distribution function $\rho_\perp(r_\perp)$, [Fig.~\ref{fig:rhoL}(a)], confirms the 
local translational order within the layers. It shows maxima at $r_\perp/\bar{d} = 1, 2, \dots$,
indicating correlations between molecules within the same layer. However, these correlations phase out
for $r_\perp/\bar{d} > 9$, indicating that the translational order is only local within the layers. 

The correlation function $S^{110}_\perp(r_\perp)$ [Fig.~\ref{fig:S110}(b)] shows 
no long-range correlation of molecular polarization vectors in SmA.
For $r_\perp/\bar{d} \approx 1$ a slight anticorrelation [$S^{110}_\perp(r_\perp) < 0$] is visible due to 
entropic reasons, where nearest-neighbour molecules tend to orient in opposite ways 
to increase packing density. However, these correlations vanish for $r_\perp/\bar{d} > 4$.
Similarly, the splay correlations [Fig.~\ref{fig:angle}(b)] remain only local within the smectic A phase. 
The preferred angle for splay correlations is lower ($\theta \approx 12^\circ$) compared to N phase, 
and the spread of angles is also lower, which is consistent with the higher value of the nematic order 
parameter $\expval{P_2}$ in the smectic phase.

Both sequences of phase transitions displayed by the system
\begin{enumerate}
    \item Iso $\leftrightarrow$ N $\leftrightarrow$ SmA
    \item Iso $\leftrightarrow$ SmA
\end{enumerate}
are prominent and well recognized in systems of elongated molecules, both in 
computational studies \cite{Vega2001,McGrother1996,Chiappini2021,Kubala2022}, 
and in experiments \cite{Mukherjee2021, Singh2000, Kumar2017}, although the second 
sequence is less common in physical systems. 

For asymmetric molecules, modulated and/or polar liquid crystalline phases may form (see the Introduction section). 
However, in our model, such phases were not observed.
The absence of ferroelectric long-range order, even in the metastable regime, is further 
confirmed by bifurcation analysis using Parsons-Lee Density Functional Theory (see Section~\ref{sec:bifurcations}). 
Two possible reasons for this absence can be considered:
the first one is the moderate length of the molecule, which may hinder the formation of long-range 
ferroelectric-like order 
and the second one is the concavities between the beads in the molecular structure.
We, however, note that preliminary studies of analogous systems of wedge-shaped molecules built of up to eleven beads as well as ones with the smooth, convex surface show no considerable change in the predictions presented by the current study.

\subsection{Crystalline phases}

Over $\eta \approx 0.5$, four distinct types of hexagonally ordered solids that retain
the SmA layered structure appear. 

\subsubsection*{Hexagonal non-polar crystal $\text{Cr}_\text{hex}$}

For $d \in [0.75, 1.0]$, SmA is adjacent to a 
hexagonal non-polar crystal phase denoted $\text{Cr}_\text{hex}$, whose snapshot is 
shown in Fig.~\ref{fig:pack_crystal}(a). Both $\expval{P_2}$ and $\expval{\tau}$ 
are almost equal to 1 in the entire range, as seen in [Fig.~\ref{fig:pd_obs}(a, b)]. 
Additionally, there is a sharp jump in the local hexatic order parameter,
as depicted in [Fig.~\ref{fig:pd_obs}(c)], confirming the presence of a locally 
hexatic structure within a layer. 
The long-range translational order becomes apparent when observing the radial and 
layer-wise pair distribution functions $\rho(r)$ and $\rho_\perp(r_\perp)$, respectively,
as shown in [Figs.~\ref{fig:rho}(d),\ref{fig:rhoL}(b)]. 
In the case of $\rho(r)$, we observe two sequences of maxima. 
Similar to SmA, the first sequence with larger spacing corresponds to layering, 
while the second sequence with a smaller peak-to-peak distance 
corresponds to in-layer order. However, unlike SmA, the second sequence 
does not vanish quickly and extends throughout the entire plotted 
range of $r/\bar{d}$. Similarly, $\rho_\perp(r_\perp)$ also
exhibits long-range correlations. Additionally, the structure of the first few peaks 
of $\rho_\perp(r_\perp)$, as shown in Fig.~\ref{fig:rhoL}(f), agrees with the one 
observed for the hexagonal honeycomb lattice (see eg. Fig.~3 of Ref.~\cite{Brodin2010}). 
The $S^{110}_\perp(r_\perp)$ correlations, as depicted in [Fig.~\ref{fig:S110}(c)], 
indicate weak and extremely short-range polarization order, which disappears 
for $r_\perp/\bar{d} > 2$. A similar observation can be made for the 
splay correlations shown in [Fig.~\ref{fig:angle}(c)], where the preferred angle 
is close to $0^\circ$ with a very narrow spread. The presence of a series of 
additional vertical lines is possibly a result of lattice defects.

For $d=1$, the wedge reduces to a well-known linear tangent hard-sphere 
(LTHS) hexamer. 
The equilibrium phases in the LTHS model have been studied in Ref.~\cite{Vega2001}. 
The ground state configuration in this case is a close pack arrangement of 
spheres \cite{Hales2005}, with a packing fraction of $\eta = \pi/\sqrt{18} \approx 0.74$. 
In our simulations, we were able to achieve a packing fraction of approximately 
$\eta \approx 0.73$ upon compression, which is in good agreement with the exact value, 
taking into account the presence of a small number of defects in the Monte Carlo 
configuration. There are various ways to arrange LTHS chains into a close-packed 
configuration. One obvious choice is to form face-centered cubic (fcc) or 
hexagonal close-packed (hcp) layers of molecules with a $30^\circ$ tilt 
(e.g., the $\text{CP}_1$ lattice in Ref.~\cite{Vega1992}, also utilized 
in the previously mentioned Ref.~\cite{Vega2001}). However, these configurations 
are unlikely to form during the compression route from SmA layers.

More probable are non-tilted variants with a partial hexatic order. 
We have identified two such structures referred to as $\text{fcc}_\text{B}$ 
and $\text{hcp}_\text{B}$ (see Appendix~\ref{sec:hexagonal}).
In these structures, six nearest neighbors within a layer form slightly deformed regular 
hexagons. This deformation is consistent with the  values of $\expval{\psi_6}$ 
below 0.8 for $d=1$, compared to over 0.8 for $d=0.85$. 
In the $\text{fcc}_\text{B}$ and $\text{hcp}_\text{B}$ lattices, 
the polymers arrange themselves into infinite columns that can be 
translated by integer multiples of the diameter d = 1 of the monomers. 
Our results exhibit a slight spread off-layer
[observed in the last column of Fig.~\ref{fig:pack_crystal}(a)], 
which also contributes to lowering the value of $\expval{\tau}$.

\subsubsection*{Antiferroelectric splay crystal $\text{Cr}_\text{S}\text{P}_\text{A}$}

When $d$ is decreased and the variation of bead diameters increases, 
the preference for a closed-packed configuration diminishes, 
and crystalline polar blue phases emerge. 
The first one is visible in a range around $d =$ 0.85-0.95 (see Fig.~\ref{fig:pd}). 
The corresponding snapshot is presented in Fig.~\ref{fig:pack_crystal}(b). 
As clearly visible in the snapshot, within each layer, 
clusters with macroscopic polarization spontaneously emerge and arrange themselves
in a striped metastructure. Adjacent clusters exhibit opposite polarization 
and are separated by planar defects in the polarization field. 

Within a cluster, the molecules have a long-range translational order, which is
indicated by the correlation function $\rho_\perp(r_\perp)$ [Fig.~\ref{fig:rhoL}(e)] and they form a hexagonal structure, reflected in a high value of $\expval{\psi_6}$ [Fig.~\ref{fig:pd_obs}(c)].
The density correlations exhibit a complex structure. 
Firstly, there is a series of wide maxima and minima separated 
by $\Delta r_\perp \approx 6\bar{d}$, which corresponds to the width of the stripes. 
Around these wide maxima, there are 3-4 sharper ones 
separated by $\Delta r_\perp \approx \bar{d}$, representing 
correlations between particular rows of shapes across the stripes. 
Additionally, there are numerous small maxima, representing correlations 
between individual beads in a highly regular lattice. However, this phase appears only 
for very high $\eta$ values close to the maximal packing. The macroscopic 
polarization of the cluster is confirmed by $S^{110}_\perp(r_\perp)$ correlations 
[see Fig.~\ref{fig:S110}(d)]. Similarly, for $\rho_\perp(r_\perp)$, we observe 
minima and maxima, but with a two times smaller separation, as the adjacent stripes 
have opposite polarizations. Furthermore, there is a similar series of additional 
maxima as observed in $\rho_\perp(r_\perp)$.

The nematic order $\expval{P_2}$ remains above 0.8 
but is slightly lower than for $\text{Cr}_\text{hex}$. This can be explained 
by the observation that clusters exhibit a splay modulation in the polarization field.
Specifically, the polarization is parallel to the director in the middle of the stripe 
and gradually leans when one moves towards the boundary. However, the preferred direction 
does not change along the length of a stripe. This observation is supported 
by the $P(\theta|r_\perp)$ histogram [Fig.~\ref{fig:angle}(d)], where, for the first time, 
a linear growth of the preferred angle $\theta$ with $r_\perp$ is observed 
until $r_\perp \approx 7 \bar{d}$. After this point, a linear decline follows. 
The ascending part of the histogram corresponds to correlations within a single stripe, 
while the falling part corresponds to correlations between adjacent stripes. 

Entropically promoted pure splay, inherently related to the shape 
of the wedge, cannot be extended globally without 
introducing energetically expensive defects. The observed two dimensional splay modulation 
is a way out of this difficulty (frustration), allowing for a more efficient
filling of space.
Notably, the pattern of clusters in adjacent layers continues, but their polarization always 
has an opposite sign. The same sign reversal applies to the direction of 
splay modulation. Consequently, 
the stripes ($y$-axis columns) are arranged in a two-dimensional checkerboard pattern. 
The largest and smallest balls that form the molecules create hexagonal lattices 
with different lattice constants. The polarization switch 
in a neighboring layer facilitates a more compatible arrangement of adjacent layers. 

To capture the most important properties of the structure, 
we refer to it as the \textit{antiferroelectric splay crystal}, 
denoted as  $\text{Cr}_\text{S}\text{P}_\text{A}$. The term \textit{antiferroelectric} 
refers to opposite polarization of the clusters in adjacent layers. It is worth noting 
that the $\text{Cr}_\text{S}\text{P}_\text{A}$ phase emerges even when the value of
$d$ is close to $1$. However, the phase boundary significantly increases as $d$ grows.
This behavior can be attributed to the requirement of higher densities to 
induce polar order when the gradient of ball diameters is smaller.
In the entire range, the system first crystallizes
into the $\text{Cr}_\text{hex}$ structure, followed by a transition to the
$\text{Cr}_\text{S}\text{P}_\text{A}$ phase. Therefore, there is no direct
SmA-$\text{Cr}_\text{S}\text{P}_\text{A}$ phase transition present.

\subsubsection*{Antiferroelectric double splay crystal $\text{Cr}_\text{DS}\text{P}_\text{A}$}

Above $\text{Cr}_\text{hex}$, another polar phase is observed  for $0.75 \le d \le 0.8$. 
The corresponding snapshot is shown in Fig.~\ref{fig:pack_crystal}(c). 
In this phase, polarization clusters are formed again, but they arrange themselves 
within a layer in a checkerboard pattern instead of stripes.
Each square domain is surrounded by four domains with opposite polarization. 
When moving to an adjacent layer, the polarizations of the clusters are flipped, 
similar to $\text{Cr}_\text{S}\text{P}_\text{A}$.
This indicates that the checkerboard pattern is actually three dimensional. 

The splay pattern also changes in this phase. It becomes three-dimensional, 
where the direction of the splay vector
$\vu{n} (\div{\vu{n}})$ is correlated with the polarization field. 
The molecules within the cluster are parallel to the z-axis in the middle 
and gradually lean as the distance from the center increases. This structure 
is referred to as the \textit{antiferroelectric double splay crystal} 
($\text{Cr}_\text{DS}\text{P}_\text{A}$). 
Importantly,  the magnitude of the splay vector in the double splay structure is higher
compared to that in the single splay structure (assuming the same maximal curvature). 
This higher splay deformation is facilitated by molecules with smaller values of $d$,
which promote a stronger modulation of the director field 
ultimately leading to the formation of the checkerboard pattern observed 
in $\text{Cr}_\text{DS}\text{P}_\text{A}$.
It is also reflected in a slightly lower value of $\expval{P_2}$ compared 
to $\text{Cr}_\text{S}\text{P}_\text{A}$. 

In the $\text{Cr}_\text{DS}\text{P}_\text{A}$ phase, the transversal density correlation
$\rho_\perp(r_\perp)$ [Fig.~\ref{fig:rhoL}(c)] exhibits a similar behavior as in the 
$\text{Cr}_\text{hex}$ [Fig.~\ref{fig:rhoL}(b)], but with higher damping. 
The angle histogram $P(\theta|r_\perp)$ [Fig.~\ref{fig:angle}(e)] in the
$\text{Cr}_\text{DS}\text{P}_\text{A}$ phase
is also similar to that in the $\text{Cr}_\text{S}\text{P}_\text{A}$ phase [Fig.~\ref{fig:angle}(d)],
but with larger preferred angles. Again, the difference is a result 
of stronger splay deformation in  
$\text{Cr}_\text{DS}\text{P}_\text{A}$ compared to 
$\text{Cr}_\text{S}\text{P}_\text{A}$.
Regarding the polarization correlation function $S_\perp^{110}(r_\perp)$ 
[Fig.~\ref{fig:S110}(e)] it shows positive correlations up to $r_\perp \approx 6.5\bar{d}$, 
and then negative correlations up to $r_\perp \approx 13\bar{d}$. 
This behavior is consistent with the snapshot in Fig.~\ref{fig:pack_crystal}(c), 
where the cluster radius is estimated to be approximately (6-7)
times the average diameter  of balls that
build the molecule (6-7$\bar{d}$).

\subsubsection*{Ferroelectric double splay crystal $\text{Cr}_\text{DS}\text{P}_\text{F}$}

For $d < 0.75$, the third frustrated polar structure 
called the \textit{ferroelectric double splay crystal} ($\text{Cr}_\text{DS}\text{P}_\text{F}$)
is formed directly above the  SmA phase. 
Similar to $\text{Cr}_\text{DS}\text{P}_\text{A}$, the $\text{Cr}_\text{DS}\text{P}_\text{F}$ 
phase also exhibits the formation 
of polar clusters arranged in a checkerboard pattern [\textit{cf.} Fig.~\ref{fig:pack_crystal}(d)]. 
However, unlike $\text{Cr}_\text{DS}\text{P}_\text{A}$ 
where the clusters alternate between neighboring layers,
in $\text{Cr}_\text{DS}\text{P}_\text{F}$ the clusters  
form vertical columns along the $z$-axis that extend throughout 
the entire height of the simulation box.

The formation of these columns is accompanied by the bending of layers, as seen 
in the second and third panels of Fig.~\ref{fig:pack_crystal}(d). Additionally, the 
decrease in the smectic order parameter $\expval{\tau}$ to the range of 0.4-0.7
further indicates on a curved layer structure. 
These curved layers cause the hexatic arrangement to deform when orthogonally projected, 
resulting in a lower hexatic order parameter $\expval{\psi_6}$ in the range of 0.65-0.75. 
The weaker nematic order $\expval{P_2}$ in the range of 0.5-0.8 is also consistent 
with the observed curved layers, which induce significant splay modulation.

Why is the $\text{Cr}_\text{DS}\text{P}_\text{F}$ structure stabilized for $d < 0.75$?
For a simple explanation, please note that
in $\text{Cr}_\text{S}\text{P}_\text{A}$ and $\text{Cr}_\text{DS}\text{P}_\text{A}$, the opposite 
interlayer polarization allows for the adjacency of compatible hexagonal bead lattices.
However, this arrangement enforces flat layers, which becomes inefficient for smaller values of  $d$.
An efficient packing arrangement that leads to higher packing densities
involves clusters that resemble fragments of a spherical shell.
Achieving such an arrangement would require a ferroelectric alignment in the direction of 
the vector $\vb{k}$ to match the signs of curvatures. 
Interestingly, this is precisely what happens  
in the transition from $\text{Cr}_\text{DS}\text{P}_\text{A}$ to $\text{Cr}_\text{DS}\text{P}_\text{F}$. 
The formation of columns in $\text{Cr}_\text{DS}\text{P}_\text{F}$ allows for layer bending 
and the emergence of a mutually coupled ferroelectric arrangement in this phase. 
Although  this ordering sacrifices the 
compatibility of the lattice constants of the beads, the packing microstructurure 
remains similar to 
$\text{Cr}_\text{DS}\text{P}_\text{A}$.

There are long-range correlations observed in the transversal density correlation
$\rho_\perp(r_\perp)$ [Fig.~\ref{fig:rhoL}(d)], which, however, vanish 
faster than for $\text{Cr}_\text{DS}\text{P}_\text{A}$ or $\text{Cr}_\text{hex}$. 
This can be attributed to more irregular domain walls, which weaken the correlations 
between adjacent clusters. 
The polarization correlation function
$S^{110}_\perp(r_\perp)$ [Fig.~\ref{fig:S110}(f)] in $\text{Cr}_\text{DS}\text{P}_\text{F}$
is similar to that in $\text{Cr}_\text{DS}\text{P}_\text{A}$, and the cluster radius estimated at 
7-8 molecule diameters agrees with a visual inspection of Fig.~\ref{fig:pack_crystal}(d). 
The preferred angles on the angle histogram $P(\theta|r_\perp)$ [Fig.~\ref{fig:angle}(f)] 
grow linearly with $r_\perp$ confirming the presence of long-range splay correlations. 
However, unlike in Fig.~\ref{fig:angle}(d,e), a linear descent in the angle histogram is not observed in 
$\text{Cr}_\text{DS}\text{P}_\text{F}$.

Finally, with decreasing $d$ the variation of ball diameters grows larger, 
which hinders the optimal filling of the space. As a result, 
estimated maximal packing fraction $\eta$ (see black solid line in Fig.~\ref{fig:pd}) 
falls monotonically with $d$ reaching $\eta \approx 0.61$ for $d = 0.4$.

\section{Density functional analysis of  
liquid crystalline order} \label{sec:bifurcations}
Simulations show that the shape-induced splay deformations in the system 
of wedge-shaped hard molecules can lead to unusually complex long-range orientational 
ordering in crystalline phases. What is somewhat unexpected  
is that the observed long-range orientational order in 
the liquid crystalline phases is not affected much. 
That is, only the ordinary uniaxial nematic and smectic A phases are found at equilibrium.
The question remains of whether liquid crystalline phases of non-trivial orientational 
ordering, such as \textit{e.g.} ferroelectric nematic/smectic A, 
antiferroelectric smectic A or splay nematic/smectic, can form 
subject to purely steric interactions. 
Although we did not observe these in simulations as equilibrium states, 
it is imperative to assess whether such polar liquids be formed at least in a metastable regime.
The simplest way to approach this issue is to perform density 
functional bifurcation analysis, in which the form of stable/metastable 
states can be selected at the start. 
In the following, we concentrate on the potential (meta)stability 
of ferroelectric nematic ($\text{N}_\text{F}$), smectic A ($\text{SmA}$) and (anti)ferroelectric 
smectic A ($\text{SmA}_\text{AF}$)
using the second-virial Density Functional Theory corrected by the Parsons-Lee term 
(DFTPL)\cite{Onsager1949effects,parsons1979nematic,lee1987numerical}.
We will seek to gain a better understanding  
of the ability of wedge-shaped molecules to form 
liquid phases with polar order. 
The study of the $\text{N}-\text{SmA}$ bifurcation here serves as a reference.

The use of the DFTPL approach proved especially useful in the analysis 
of (meta)stable structures for hard molecules
of complex shapes (see \textit{e.g.} \cite{GrecoFerrariniPRL}).
We start with a brief summary of the DFTPL 
theory for hard uniaxial molecules. More details can be found \textit{e.g.}
in \cite{GrecoFerrariniPRL, Longa2005,  Karbowniczek2017}.  According to this theory, 
the Helmholtz free energy $\mathcal{F}$ of a system of $N$ molecules in volume $V$ 
at temperature $T$ is a functional of the single molecule probability density distribution
function $P(X_i)$, where $X_i\equiv \{\mathbf{x}_i,\mathbf{\hat{a}}_i \}$ 
represents the position and orientation of the i\textit{-th} molecule. The distribution 
$P(X_i)$ is normalized so that 
\begin{equation}\label{normalization}
 \Tr_{(X_i)}P(X_i)=1, \hspace{0.5cm}    \Tr_{(X_i)}=\int_{V}d^3\mathbf{r}_i\int 
 d^2 \mathbf{\hat{a}}
\end{equation}
After disregarding terms that can be made independent of $P$, the relevant part of 
the free energy per molecule for hard uniaxial molecules of arbitrary shape can be written as
\begin{eqnarray}
\label{eq:freeenergy}
 f(P)\equiv  \frac{ \mathcal{F}[P] }{N k_\text{B} T}  & = &  \underset{(X)}{
\Tr}\left[ P(X)\ln  P(X) \right] \nonumber\\ && \hspace{-2cm}+\, C(\eta)\frac{
{\bar{\rho }}}{2}\,\underset{(X)}{
\Tr}\left[ P(X)H_\text{eff}(X,[P])\right] + ...\, ,
\end{eqnarray}
where $\bar{\rho}=\frac{N}{V}$ is the average density.  
$H_\text{eff}$ is the effective excluded volume 
averaged over the probability distribution of molecule ``2'' ($X \equiv X_1$):
\begin{equation}
H_\text{eff}(X_{1},[P])= V \underset{(X_{2})}{
\Tr}\left\{
\Theta \left[ \xi (X_{1},X_{2})-r_{12}\right]\, P(X_{2}) \right\},
\label{Heff13}
\end{equation}
where 
$\xi(\cdot)$ is the contact distance between two molecules, 
$\Theta(\cdot)$ is the Heaviside $\Theta$ function and $k_B$ is the Boltzmann constant. 
The factor $C(\eta)=(1/4)\frac{4-3\eta}{(1-\eta)^2}$
renormalizes the second-order virial expansion \cite{Onsager1949effects} to take into account
the higher-order terms of the expansion \cite{parsons1979nematic,lee1987numerical}. Finally,
$r_{12}$ is the distance between two molecules,
$\eta=\frac{v_\text{mol}N}{V}=v_\text{mol}\bar{\rho}$ is the packing fraction,
and $v_\text{mol}$ is the volume of a molecule.

The equilibrium states correspond to the minimum of the free energy 
functional, Eq.~(\ref{eq:freeenergy}), 
with respect to $P(X)$, subject to the normalization condition, Eq.~(\ref{normalization}).
The procedure is equivalent to solving the self-consistent integral equation 
for the stationary distributions $P_s(X)$:
\begin{equation}
\label{eq:selfconsistent}
    P_s(X) = Z^{-1}\exp[-\bar{\rho} C(\eta)H_\text{eff}(X,[P_s])]
\end{equation}
where 
\begin{equation}
    Z = \underset{(X)}{
\Tr} \exp[-\bar{\rho} C(\eta) H_\text{eff}(X,[P_s])]
\end{equation}
and selecting the solution that minimizes the free energy (\ref{eq:freeenergy})
for a given set of control parameters. In general, for wedge-shaped molecules, it is important to know  
whether they can stabilize states with some kind of polar order,
such as polar nematics or smectics. The effective method for exploring this problem
is the bifurcation analysis of Eq.~(\ref{eq:selfconsistent}) about the reference
state, usually the isotropic or uniaxial nematic phase.   
Since wedges orient their steric dipole, on average, along the 
director $\mathbf{\hat{n}}(\mathbf{r})$, 
different local polar order of the  
predefined polarization profile $\mathbf{\hat{p}}(\mathbf{r})$ ($|\mathbf{\hat{p}}(\mathbf{r})|=1$,
$\mathbf{\hat{p}}(\mathbf{r})=\pm\mathbf{\hat{n}}(\mathbf{r})$) can be tested
against the instability of the reference state.

Here, we focus on bifurcation studies from the reference uniaxial nematic phase.   
As simulations show, the nematic order is high in both 
the uniaxial nematic and at the 
transition from the uniaxial nematic to higher ordered phases (see Fig.~\ref{fig:pd}), 
so we can further simplify the analysis by assuming that the orientational order is 
saturated. Consequently, the orientational degrees of freedom of the $\text{i}^\text{th}$ polar molecule  
can be described by a discrete pseudospin variable, $s_i=\pm 1$. It tells whether the steric 
molecular dipole is parallel ($s_i = 1$) or antiparallel ($s_i = -1$) 
to the preferred local orientational ordering, which is assumed to be positionally independent:
\begin{equation}\label{inorder}
   \mathbf{a}_i = s_i \mathbf{\hat{p}}.
\end{equation}
In this case, the integration of orientational degrees of freedom is reduced to the sum 
over $s_i=\pm 1$, which means that 
\begin{eqnarray}\label{normalization1}
 \Tr_{(X_i)}[...]&=& \Tr_{(\mathbf{r}_i,s_i)}[...]\equiv \Tr_{(x_i,y_i,z_i,s_i)}[...] \nonumber\\
 &&
 =\int_{V}d^3\mathbf{r}_i\sum_{s_i=\pm 1}(...).
\end{eqnarray}
A more complex case of the nematic/smectic splay is left for an in-depth discussion
elsewhere.

In practical calculations, we model the stationary distribution function $P_s\equiv P(\mathbf{r},s)\equiv P(z,s)$ 
by expanding this to leading order in order parameters that describe the structures of interest.
By limiting to nematic and smectic 
polar orderings of the constant $\mathbf{\hat{p}}$ 
(parallel to the z-axis), this gives
\begin{eqnarray}\label{P-expansion}
     P(z,s)&=& \frac{1}{V}\left[
    \frac{1}{2}+ \frac{1}{2}\expval{s}s + \expval{\cos} \cos\left( \frac{2 \pi z}{d'}\right) \right.
    \nonumber\\
    +\,\,\,\,\,\, && \hspace{-0.7cm}
    \expval{\sin} \sin\left( \frac{2 \pi z}{d'}\right) +
    \expval{ s \cos}  s \cos\left( \frac{2 \pi z}{d''}\right)\nonumber\\
    &+& \left.
    \expval{ s \sin}  s \sin\left( \frac{2 \pi z}{d''}\right) + ...
    \right],
\end{eqnarray}
where
\begin{eqnarray}
    \{\expval{s}, \expval{\cos}, ..., \expval{ s \sin}\}&=& \nonumber\\
   && \hspace{-5cm}
    \Tr_{(\mathbf{r}_i,s_i)}P(z,s)\left\{
    s, \cos\left( \frac{2 \pi z}{d'}\right), ..., 
    s \sin\left( \frac{2 \pi z}{d''}\right)
    \right\}
\end{eqnarray}
are the order parameters 
and where the length of the box ($V^\frac{1}{3}$) is 
assumed to be a multiple of the smectic periods $d'$ and $d''$.

Note that in expansion, Eq.~(\ref{P-expansion}), the order parameter
$\expval{s}$ represents the long-range polar order of the molecules in the nematic, 
smectic and crystalline phases, while
$\langle s \rangle \mathbf{\hat{p}}$ is the average polarization.
In the case of $\text{SmA}$ with density modulation 
along the $z$ axis of the laboratory frame, only the order parameter $\expval{\cos}$ is not zero 
while $d'$ is the period of the structure. This phase is always stable in simulations 
and will therefore serve as a test for bifurcation theory. 
The remaining order parameters can be combined to
form antiferroelectric smectic phases, but detailed predictions depend on
the solutions of Eq.~(\ref{eq:selfconsistent}).

The calculations can now proceed by pointing out that the order parameters 
are small near the bifurcation point. 
This enables us to linearize Eq.~(\ref{eq:selfconsistent}) 
for a very small non-zero value of $\delta P(z,s)$, where $\delta P(z,s)=P(z,s)-P_0$; 
$P_0=\frac{1}{2V}$ takes into account the probability distribution of the ideally 
oriented uniaxial nematic phase. The resulting linear homogeneous equation 
for $\delta P(z,s)$ is given by
\begin{eqnarray}\label{linearized_eq}
    \delta P(z_1,s_1)&=& -\eta\, C(\eta) \left(\frac{1}{2 v_\text{mol}}\right) \times \nonumber\\
   && \hspace{-2cm}   \Tr_{({z}_{},s_2)}
    \left[
\Theta({z}_{},s_1,s_2)\, \delta P(z_{}+z_1,s_2) \right].
\end{eqnarray}
Here
 $\Theta({z},s_1,s_2)= Tr_{(x_{12},y_{12})}\Theta \left[ \xi (\mathbf{r}_{12},s_1,s_2)-r_{12}\right]$ ($z\equiv z_{12}$),
 $v_\text{mol}$ is the volume of the molecule.

Before identifying the phases that can bifurcate from $N$, we observe that the excluded 
interval $\Theta({z}_{},s_1,s_2)$ has a particularly simple form in relation to the variables $\{s_1,s_2\}$.
Specifically, observing the symmetry of $\Theta({z}_{},s_1,s_2)$: $\Theta({z}_{},-1,1)=\Theta(-{z}_{},1,-1)$
and $\Theta({z}_{},-1,-1)=\Theta({z}_{},1,1)=\Theta(-{z}_{},1,1)$ we can replace $\Theta({z}_{},s_1,s_2)$
with the sum
\begin{eqnarray}\label{decomp}
 \Theta({z}_{},s_1,s_2)=  \Theta_0({z}_{})+ s_1 s_2 \Theta_1({z}_{})
  +(s_1-s_2)\Theta_2({z}_{}), \nonumber \\
\end{eqnarray}
where
\begin{eqnarray}
 \Theta_0(z)&=& \Theta_0(-z) = \nonumber\\
 && \hspace{-1cm} \frac{1}{4}\left[
 2\, \Theta(z,1,1) + \Theta(z,1,-1)+\Theta(-z,1,-1)
 \right] \\
 \Theta_1(z)&=& \Theta_1(-z) = \nonumber\\
 && \hspace{-1cm} \frac{1}{4}\left[
 2\, \Theta(z,1,1) - \Theta(z,1,-1)-\Theta(-z,1,-1)
 \right] 
\end{eqnarray}
and where
\begin{eqnarray}
 4\Theta_2(z)= -4\Theta_2(-z) = 
 \Theta(z,1,-1)-\Theta(-z,1,-1). \nonumber \\
\end{eqnarray}
It should be noted that the term proportional to $(s_1+s_2)$ disappears due to the symmetries mentioned above
of $\Theta({z}_{},s_1,s_2)$.

Now, the homogeneous Eq.~(\ref{linearized_eq}) can be solved for $\delta P(z_1,s_1)$, and the solutions 
are parameterized by the corresponding packing fraction $\eta=\eta_b$. 
More specifically, for $P(z,s)$ given by Eq.~(\ref{P-expansion}) the Eq.~(\ref{linearized_eq})
becomes reduced to a set of homogeneous equations for the order parameters.
They are given by
\begin{eqnarray}
    \langle s \rangle &=& \langle s \rangle \psi(\eta_b) \, \Theta_{1,s}
    \label{linearized_order_parameters_for_s}\\ \nonumber\\
    \left(\!\!\!
    \begin{array}{c}
         \expval{\cos}    \\
         \expval{ s \sin} 
    \end{array}\!\!\!
    \right) &=& \psi(\eta_b)\times\nonumber\\
    && \hspace{-3cm}
    \left( 
    \begin{array}{cc}
        \Theta_{0,c}(d') & -\Theta_{2,s}(d'')\delta_{d',d''} \\
        -\Theta_{2,s}(d'')\delta_{d',d''} & \Theta_{1,c}(d'')
    \end{array}
    \right)
    \left(\!\!\!
    \begin{array}{c}
         \expval{\cos}    \\
         \expval{ s \sin}
    \end{array}\!\!\!
    \right)\nonumber\\\nonumber\\
    \left(\!\!\!
    \begin{array}{c}
         \expval{\sin}    \\
         \expval{ s \cos} 
    \end{array}\!\!\!
    \right) &=& \psi(\eta_b)\times\nonumber\\
    && \hspace{-3cm}
    \left( 
    \begin{array}{cc}
        \Theta_{0,c}(d') & \Theta_{2,s}(d'')\delta_{d',d''} \\
        \Theta_{2,s}(d'')\delta_{d',d''} & \Theta_{1,c}(d'')
    \end{array}
    \right)
    \left(\!\!\!
    \begin{array}{c}
         \expval{\sin}    \\
         \expval{ s \cos} 
    \end{array}\!\!\!
    \right)
    \label{linearized_order_parameters}
    \\ \nonumber
\end{eqnarray}
where
\begin{eqnarray}
\Theta_{1,s}&=& \int_{-l}^{l} \Theta_1(z) \mathrm{d}z\nonumber\\
  \Theta_{0,c}(d')&=&\int_l^l\Theta_0(z) \cos(\frac{2 \pi z}{d'})\nonumber\\
  \Theta_{2,s}(d') &=&\int_l^l\Theta_2(z) \sin(\frac{2 \pi z}{d'})\nonumber\\
  \Theta_{1,c}(d'') &=&\int_l^l\Theta_1(z) \cos(\frac{2 \pi z}{d''})
\end{eqnarray}
and where $    \psi=-\frac{2 \pi}{v_\text{mol}}\eta C(\eta) $;  
$l$ is the molecular length. We should add that with Wolfram Mathematica
the formulas for the coefficients $\Theta_{\alpha,\beta}$ can be found exactly for 
rational diameters of the spheres.

\textit{A priori} one expects four types of bifurcating states from Eq.~(\ref{linearized_order_parameters}). 
The first is the ferroelectric phase ($F$), where only $\langle s \rangle $ 
becomes nonzero at $\eta_b$. Due to the symmetry of the reference state $P_0$
and of the excluded interval, Eq.~(\ref{decomp}), first-order bifurcation analysis 
does not lead to a coupling between $\expval{s}$ and the 
smectic or crystalline order. Thus, if any $F$
results from Eq.~(\ref{linearized_order_parameters}), it cannot be fully identified 
and can actually correspond to
a ferroelectric order of a nematic, smectic, or crystal phase.
To resolve which of the cases applies, a higher-order bifurcation analysis is needed
in this case. For $d'\ne d''$ we expect classical smectic A ($\text{SmA}$) with nonzero $\langle \cos \rangle$
(equivalently $\langle \sin \rangle\ne 0 $) and antiferroelectric smectic A 
($\text{SmA}_\text{AF}$) where $\langle  s \cos\rangle \ne 0$
(equivalently $\langle  s \sin \rangle \ne 0$). The final possibility is where $d'= d''$.
In this case, we expect the $\text{SmA}_\text{d}$ phase, where $\langle \cos \rangle\ne 0$ and $\langle  s \sin \rangle \ne 0$,
(equivalently  $\langle \sin \rangle\ne 0$ and $\langle  s \cos \rangle \ne 0$). The phase would be 
similar to $\text{SmA}$, but with $d'$ incommensurate with $l$. As two order parameters condense at the bifurcation to
$\text{SmA}_\text{d}$ the corresponding phase transition should generally be of the first order.

The structure to stabilize 
as a result of the phase transition from uniaxial nematic 
is usually (but not always) the one that leads to the minimum value of $\eta_b$. In the case of smectics, 
the bifurcation packing fraction $\eta_b$ also depends on $d'$,  
which requires additional minimization of $\eta_b$ with respect to the smectic period.
The hierarchy of $\eta_b$-s gives an idea of possible (meta)stable states that the model 
can predict.

We begin our detailed analysis by determining 
whether any type of long-range polar order can occur in our model. 
The solution $\langle s \rangle \ne 0$ bifurcates from N
at $\eta_b$ satisfying the equation [see Eq.~(\ref{linearized_order_parameters_for_s})]
\begin{equation}\label{eta_bif_N}
    1 =  \psi(\eta_b)\, \Theta_{1,s}.
\end{equation}
Only the solution with $\eta_b>0$, where $\eta_b$ is smaller
than the maximal packing fraction, corresponds to a physically acceptable 
ferroelectric state. Clearly,  it satisfies
the integral equation (\ref{eq:selfconsistent}) for $\eta > \eta_b$. 
Note that for $\Theta_{1,s}>0$ no physical solution of Eq.~(\ref{eta_bif_N})
for $\eta_b$ exists. In this case, the excluded volume of the parallel arrangement
of the steric dipoles prevails that of the antiferroelectric one    
suggesting that the preferred local ordering should be of an antiferroelectric 
type.  The calculations reveal that for 
all $d$-parameters studied, the bifurcating packing fraction is always negative ($ \eta_b <0$), which 
means that the ferroelectric polar order is globally unstable at the expense 
of some kind of antiferroelectric ordering. Interestingly, the same conclusions can be drawn
for model molecules composed of one sphere of diameter 1 and five spheres with their diameter 
chosen at random between 
1 and 0.4. We have checked this for a sample of about 10000 different molecules.
Overall, these results suggest that, within the assumptions and simplifications 
adopted, the density functional theory does not predict the existence of global 
polar ordering in the hard model systems built out of six spheres. 

A similar analysis can be performed to study the bifurcation to $\text{SmA}$ and $\text{SmA}_\text{AF}$.
Here, the bifurcation equations ($d'\ne d''$), analogous to  Eq.~(\ref{eta_bif_N}, are given by
\begin{eqnarray}
  1 &=&  \psi(\eta_b)\, \Theta_{0,c}(d') \hspace{1cm} \mathrm{for}\,\, \text{N}-\text{SmA} 
  \label{eta_bif_SA}\\
  1 &=&  \psi(\eta_b)\, \Theta_{1,c}(d'') \hspace{1cm} \mathrm{for}\,\, \text{N}-\text{SmA}_\text{AF}
   \label{eta_bif_SAF}. 
\end{eqnarray}

A more complex case of $\text{SmA}_\text{d}$ ($d'=d''$) requires diagonalization of 
the symmetric matrix $2 \times 2$ 
\begin{eqnarray}\label{SAd_case}
    \left( 
    \begin{array}{cc}
        \Theta_{0,c}(d') & \Theta_{2,s}(d') \\
        \Theta_{2,s}(d') & \Theta_{1,c}(d')
    \end{array}
    \right).
\end{eqnarray}
It allows us to reduce the matrix equations in Eq.~(\ref{linearized_order_parameters}) to
 independent linear equations.
For example, taking the first of equations (\ref{linearized_order_parameters})
we obtain two independent linear relations similar to (\ref{linearized_order_parameters_for_s}), 
where $\Theta_{1,s}$ is replaced by one of the eigenvalues of the matrix, Eq.~(\ref{SAd_case}),
and $\expval{s}$ by a linear combination of order parameters: 
$\langle \cos\rangle + \frac{o_{12}}{o_{11}} \langle  s \sin\rangle$; $o_{ij}$-s 
are elements of the orthogonal matrix $\mathbf{o}$ 
that brings (\ref{SAd_case}) into the diagonal form. 
By inspecting Eqs.~(\ref{linearized_order_parameters},\ref{SAd_case})
we find that in our case the corresponding bifurcation equation along 
with the bifurcating state becomes
\begin{eqnarray} \hspace{-5mm}
 \frac{2}{\psi(\eta_b)} &=& 
    \Theta_{0,c}+\Theta_{1,c}-\sqrt{4 \Theta_{2,s}^2 + (\Theta_{0,c}-\Theta_{1,c})^2}\label{eta_bif_SAd}
    \hspace{6mm}
    \\ 
   \frac
   { \delta P(z,s)}{\varepsilon}  
   &=& 
    \langle \cos \rangle +  \nonumber\\
    &&\hspace{-1.5cm} 
    \frac{2 \Theta_{2,s}}{\Theta_{1,c}-\Theta_{0,c}+
    \sqrt{4 \Theta_{2,s}^2 + (\Theta_{0,c}-\Theta_{1,c})^2}} \langle s \sin \rangle, 
\end{eqnarray}
where $\varepsilon$ is an arbitrary parameter.
As previously, the physical solution is one that leads to a 
minimum of $\eta_b>0$ with respect to $d'$.

\begin{figure}[htb]
    \centering
    \includegraphics[width=0.9\linewidth]{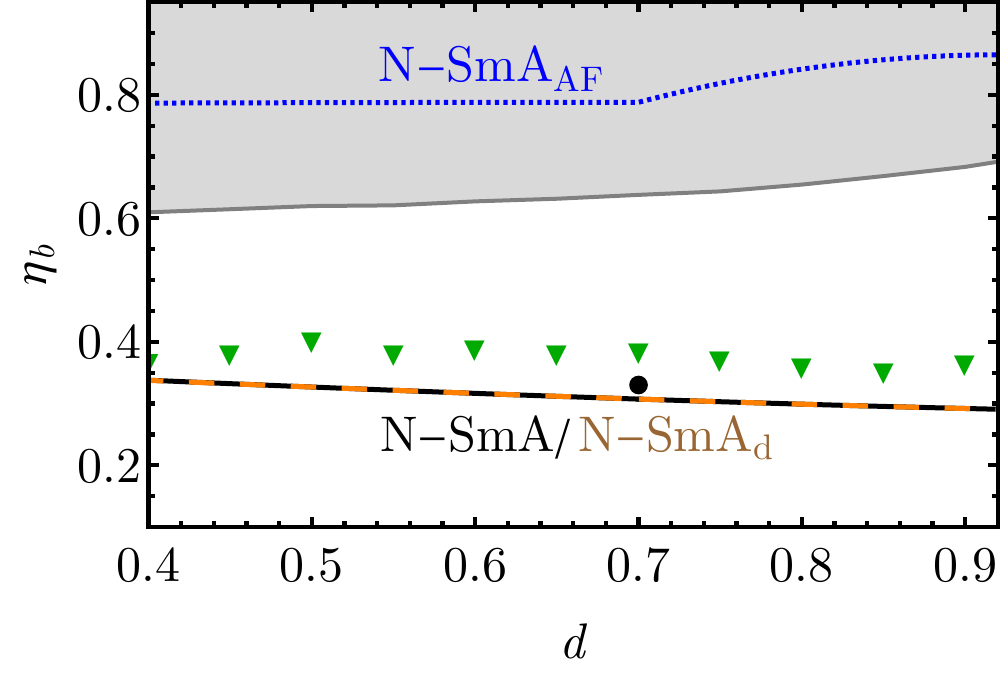}
    \caption{Bifurcations from ideally oriented nematic phase. Black-orange dashed line is two practically overlapping N-SmA and N-$\text{SmA}_\text{d}$ bifurcation lines, blue dotted line is N-$\text{SmA}_\text{AF}$ bifurcation line, and gray area is an inaccessible region of packing fractions above the optimal packing. Moreover, a black point is the N-SmA transition packing fraction $\eta \approx 0.33$ for $d=0.7$ obtained from numerical simulation with axis-aligned molecules, while green triangles are N-SmA transition packing fractions from simulations without restricting molecule orientations (cf. Fig.~\ref{fig:pd}).}
    \label{fig:bif_wedges}
\end{figure}

In Fig.~\ref{fig:bif_wedges} shown are $\eta_b$-s found 
by numerically solving the Eqs.(\ref{eta_bif_SA}, \ref{eta_bif_SAF}, \ref{eta_bif_SAd}).
Out of the assumed model structures, the one that bifurcates first is 
$\text{SmA}_\text{d}$ (continuous orange line in Fig.~\ref{fig:bif_wedges}. 
It differs from $\text{SmA}$ (black line in Fig.~\ref{fig:bif_wedges}, 
characterized by $\langle \cos\rangle$, by the presence of the
extra term $\frac{o_{12}}{o_{11}} \langle  s \sin\rangle$ that accounts for the 
polarization wave. However, the relative importance of this last term is of the order of 1\%
of the leading smectic term. The reason for detecting only $\text{N}-\text{SmA}$ in simulations is
probably the nature of the $\text{N}-\text{SmA}_\text{d}$ transition, which should generally be 
first-order due to the simultaneous condensation of two order parameters, unlike $\text{N}-\text{SmA}$. 
When comparing the simulation results with the $\text{N}-\text{SmA}$ bifurcation, we find that 
the packing fraction of the bifurcation analysis is always 
lower than predicted by the simulations.
This is due to the underestimation of the 
orientational entropy by the ideal nematic order, 
as opposed to the full spectrum of orientational degrees of freedom present
in simulations. However, if an ideal nematic order is also assumed in the simulations, 
a very good agreement between simulation and theory for $\text{N}-\text{SmA}$ is obtained  
[black dot  shown in Fig.~\ref{fig:bif_wedges}]. A good agreement is obtained  from simulations without restricting
molecule orientations [green triangles shown in Fig.~\ref{fig:bif_wedges}].
A similar analysis for $\text{N}-\text{SmA}_\text{AF}$ shows that  
$ \eta_b$ is generally of the order of $0.8$, which exceeds the physically accessible packing 
fractions for our systems.

\section{Summary}

One of the most important discoveries in the field of liquid crystals in recent years 
is the identification of the ferroelectric nematic phase (N$_\text{F}$) and the nematic phase 
with periodic long-range splay order (N$_\text{S}$). These phases were first observed in 
the RM734 molecular system.
While there are already numerous systems known to exhibit stable polar nematic phases, 
the key features of molecular interactions responsible for their stability are still under
intense studies. From the observations of molecular self-organization in the RM734 system, 
two factors appear to play a major role in stabilizing these phases. 
Specifically, the cone-like symmetry of the elongated molecule and the significantly large net 
axial dipole moment of the molecule, exceeding 10 Debye units, seem to be crucial.
This research represents a systematic effort to uncover the essential features 
of molecular selforganization that can be attributed to molecular 
shape asymetry represented by RM734. 
We focused on a molecular system composed of hard wedge-shaped 
molecules consisting of tangent spheres, where the molecular symmetry 
is controlled by a parameter called $d$, representing 
the ratio between the smallest and largest diameters of the spheres. 
An analogous hard-sphere model was previously investigated 
by Greco and Ferrarini \cite{GrecoFerrariniPRL} and some of us \cite{Kubala2022}. 
It involved modeling of bend-core-like mesogens by hard crescent-like molecules 
composed of identical beads 
that, through purely entropic interactions, stabilized another remarkable nematogenic phase, 
namely the twist-bend nematic phase. Our purpose was to investigate the type of long-range 
orientational order stabilized by molecular systems that exhibit similarity in molecular 
asymmetry to RM734. Similar to the works \cite{GrecoFerrariniPRL,Kubala2022}, we focused 
on examining the role that packing entropy can play in stabilizing such order.

Using Monte Carlo (MC) simulation, we computationally obtained and analyzed self-organization
for hard wedge-shaped molecules consisting of six tangent spheres. We simulated a wide range of packing densities, ranging from those observed 
in ordinary liquids to the maximum achievable packing fractions for a given $d$ value. 
These maximal packings are represented by continuous lines in Figures \ref{fig:pd} and \ref{fig:pd_obs}.

More systematically, for packing fractions below $\eta \approx 0.5$, which correspond to the liquid phase, 
we observed isotropic (Iso), nematic (N), and smectic A (SmA) phases, as expected for systems 
built with calamitic molecules. However, we did not observe any polar or splay nematic/smectic phases. 
This observation was further supported by Density
Functional Theory (DFT) calculations and remains valid even in the metastable regime. 
Moreover, the DFT study suggests that the polar nematic phase is unstable not only for our system 
but also for other similar systems composed of six tangent balls with a nonzero steric dipole.
Our preliminary MC simulations also indicate that 
even for analogous molecules consisting of up to eleven tangent beads or molecules with a smooth wedge surface,
there are no significant changes in the predictions. The absence of sterically induced ferroelectric 
long-range order still persists in these systems.

While a polar smectic A phase is theoretically possible, it would require unphysically high packing fractions, 
around $\eta \approx 0.8$. A noteworthy theoretical prediction involves the potential stabilization of
a bilayer smectic phase ($\text{SmA}_\text{d}$) at practically the same packing densities as those characterizing 
SmA. A similar mesophase was observed in simulations of single-site hard pears \cite{ Barmes2003}, but not in 
our simulations for wedges.

The most striking observation was the presence of complex orientational periodic superstructures, 
involving hundreds of molecules as depicted in Fig.~\ref{fig:pack_crystal},
that couple to the underlying crystalline order of molecular centers at high packing fractions. 
We label these phases as crystalline polar blue phases.
Specifically, for $\eta$ greater than 0.5, the system crystallizes 
retaining the SmA layered structure. Entropically promoted splay, inherently related
to the shape of the wedge, competes here with the tendency of layers to stay flat.
The efficient filling of space is being dependent on actual value of the parameter $d$.

For $d \ge 0.75$, when the molecule is not 
far from the linear tangent hard sphere (LTHS) hexamer, SmA transforms to a standard non-polar 
hexagonal crystal ($\text{Cr}_\text{hex}$).
However, for all $0.75 \le d < 1$ further compression 
produces two another crystalline phases that have a 
(frustrated) polar order and periodic splay modulation.
For $d$ around 0.85-0.95 (see Fig.~\ref{fig:pd}) the striped 
antiferroelectric splay crystal ($\text{Cr}_\text{S}\text{P}_\text{A}$) phase emerges.
Within each layer, clusters with macroscopic polarization are spontaneously formed,
but in the adjacent clusters the polarization is of opposite sign. The clusters 
are separated by planar defects in the polarization field.
For lower values of the $d$ parameter ($0.75 \le d \le 0.8$) the 
periodic polar stripes appear that are arranged in a 
checkerboard mesostructure. This structure is called the antiferroelectric 
double splay crystal ($\text{Cr}_\text{DS}\text{P}_\text{A}$).
Again, the the alternating polarization pattern facilitates a more efficient molecular packing.

While the splay pattern, coupled with a slight layer deformation, also changes in these two phases,
the stabilization of the third phase, called ferroelectric 
double splay crystal ($\text{Cr}_\text{DS}\text{P}_\text{F}$),
relies entirely on significant splay
modulation, which leads to strongly curved layers. In the 
$\text{Cr}_\text{DS}\text{P}_\text{F}$ phase, stabilized  for $d < 0.75$ directly from  SmA,
the ferroelectrically polarized splayed clusters 
form long columns arranged in a checkerboard pattern.  
To our knowledge, none of these phases were previously reported, 
but theoretical arguments support the existence of related mesophases
in a liquid crystalline domain \cite{Shamid2014}.


\section*{Data availability}
The datasets generated during and/or analyzed during the current study are available from P.K. upon reasonable request.

\section*{Code availability} \label{sec:code}
The source code of an original RAMPACK simulation package used to perform Monte Carlo sampling is available at \url{https://github.com/PKua007/rampack}.

\section*{Acknowledgements}
P.K. acknowledges the support of the Ministry of Science and Higher Education (Poland) 
grant no. 0108/DIA/2020/49 and, partly, the National Science Centre in Poland grant 
no. 2021/43/B/ST3/03135. M.C. and L.L. acknowledge the support of the National Science 
Centre in Poland grant no. 2021/43/B/ST3/03135. Numerical simulations were carried out 
with the support of the Interdisciplinary Center for Mathematical and Computational 
Modeling (ICM) at the University of Warsaw under grant no. G27-8.

\appendix

\section{Non-tilted hexagonal configurations in a close packing of LTHS polymers} \label{sec:hexagonal}

\begin{figure}[htb]
    \centering
    \includegraphics[width=0.7\linewidth]{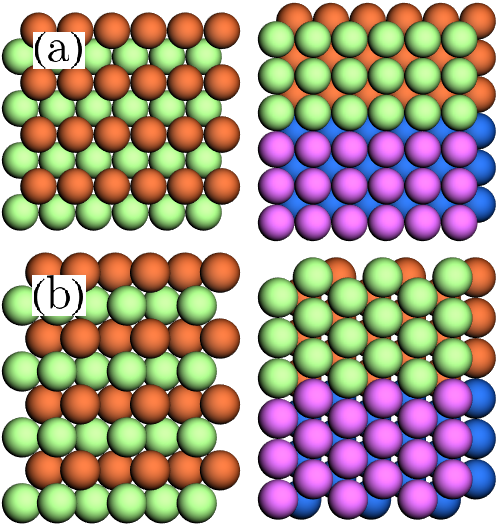}
    \caption{(a) $\text{fcc}_\text{B}$ and (b) $\text{hcp}_\text{B}$ configurations of trimers ($k = 3$). First column is a top view with a visible deformed hexatic arrangement, while the second one is a side view. Colors are to visually separate the molecules.}
    \label{fig:fccB_hcpB}
\end{figure}

We want to establish how one can arrange LTHS polymers in a maximally 
packed manner, assuming they form hexatic layers without a tilt 
(with long molecular axes perpendicular to the layers). A na\"{\i}ve 
approach would be to prepare hexagonal honeycomb layers and stack them 
like in the type B crystal \cite{Goodby2015}, which is observed 
in systems of spherocylinders \cite{Veerman1990,McGrother1996}. However, 
this configuration is not maximally packed. Therefore, we need to relax two conditions: 
molecules can deviate slightly from the layers, and hexagons can be slightly deformed. 
We have identified two such configurations based on the fcc and hcp lattices.

The first configuration is constructed from an fcc lattice of spheres [see Fig.~\ref{fig:fccB_hcpB}(a)]. 
For polymers consisting of $k$ beads with a diameter $d$, the unit cell has dimensions 
$d \times d\sqrt{2} \times kd$, 
and it contains two molecules with geometric centers: 
$(0, 0, kd/2 - d/4)$ and $(d/2, \sqrt{2}/2, kd/2 + d/4)$. 
We refer to this configuration as $\text{fcc}_\text{B}$.

The second configuration is based on the hcp lattice [see Fig.~\ref{fig:fccB_hcpB}(b)]. 
The unit cell has dimensions $d\sqrt{3} \times 2d\sqrt{6}/3 \times kd$,  and it contains 
four molecules with geometric centers: $(0, 0, kd/2 - d/4)$, $(d\sqrt{3}/3, 
d\sqrt{6}/3, kd/2 - d/4)$, $(d\sqrt{3}/2, 0, kd/2 + d/4)$ and 
$(5d/2\sqrt{3}, d\sqrt{6}/3, kd/2 + d/4)$. We refer to this structure as $\text{hcp}_\text{B}$.

\bibliography{ref}

\end{document}